\documentclass[epj,referee,nopacs]{svjour}
\usepackage{graphics}

\newcommand{\ep}{\mbox{$e^+$}}
\newcommand{\el}{\mbox{$e^-$}}

\newcommand{\pip}{\mbox{$\pi^+$}}
\newcommand{\pim}{\mbox{$\pi^-$}}
\newcommand{\pio}{\mbox{$\pi^{0}$}}
\newcommand{\fo}{\ensuremath{f_0}}

\newcommand{\wpi}{\mbox{$\omega\pi$}}
\newcommand{\mpp}{\mbox{$M_{\pi\pi}$}}
\newcommand{\mpg}{\mbox{$M_{\pi\gamma}$}}
\newcommand{\subfo}{\mbox{$_{f\!_{0}}$}}

\newcommand{\wpn}{\mbox{$\ep\el\to\omega\pio\to\pio\pio\gamma$}}
\newcommand{\dafne}{\mbox{DA$\Phi$NE}}
\newcommand{\br}{\mbox{BR}}
\newcommand{\qq}{\mbox{$q\overline{q}$}}
\newcommand{\kkbar}{\mbox{$K\overline{K}$}}
\newcommand{\qqqq}{\mbox{$qq\overline{qq}$}}
\newcommand{\lint}{\mbox{$L_{\rm int}$}}
\newcommand{\roots}{\mbox{$\sqrt{s}$}}
\newcommand{\etal}{{\it et al.}}
\title{\boldmath Dalitz plot analysis of $\ep\el\to\pio\pio\gamma$ 
  events at $\sqrt{s}\simeq M_\phi$ with the KLOE detector}

\author{The KLOE Collaboration\\
F.~Ambrosino\inst{6}, A.~Antonelli\inst{2},
M.~Antonelli\inst{2}, C.~Bacci\inst{11},
P.~Beltrame\inst{3}, G.~Bencivenni\inst{2},
S.~Bertolucci\inst{2}, C.~Bini\inst{9},
C.~Bloise\inst{2}, S.~Bocchetta\inst{11}, V.~Bocci\inst{9},
F.~Bossi\inst{2}, D.~Bowring\inst{2,13},
P.~Branchini\inst{11}, R.~Caloi\inst{9},
P.~Campana\inst{2}, G.~Capon\inst{2},
T.~Capussela\inst{6}, F.~Ceradini\inst{11},
S.~Chi\inst{2}, G.~Chiefari\inst{6},
P.~Ciambrone\inst{2}, S.~Conetti\inst{13},
E.~De~Lucia\inst{2}, A.~De~Santis\inst{9},
P.~De~Simone\inst{2}, G.~De~Zorzi\inst{9},
S.~Dell'Agnello\inst{2}, A.~Denig\inst{3},
A.~Di~Domenico\inst{9}, C.~Di~Donato\inst{6},
S.~Di~Falco\inst{7}, B.~Di~Micco\inst{11},
A.~Doria\inst{6}, M.~Dreucci\inst{2},
G.~Felici\inst{2}, A.~Ferrari\inst{2},
M.~L.~Ferrer\inst{2}, G.~Finocchiaro\inst{2},
S.~Fiore\inst{9}, C.~Forti\inst{2},
P.~Franzini\inst{9}, C.~Gatti\inst{2},
P.~Gauzzi\inst{9}, 
S.~Giovannella\inst{2,}\thanks{Corresponding author: simona.giovannella@lnf.infn.it},
E.~Gorini\inst{4}, E.~Graziani\inst{11},
M.~Incagli\inst{7}, W.~Kluge\inst{3},
V.~Kulikov\inst{5}, F.~Lacava\inst{9},
G.~Lanfranchi\inst{2}, J.~Lee-Franzini\inst{2,12},
D.~Leone\inst{3}, M.~Martini\inst{2},
P.~Massarotti\inst{6}, W.~Mei\inst{2},
S.~Meola\inst{6}, 
S.~Miscetti\inst{2,}\thanks{Corresponding author: stefano.miscetti@lnf.infn.it},
M.~Moulson\inst{2}, S.~M\"uller\inst{2},
F.~Murtas\inst{2}, M.~Napolitano\inst{6},
F.~Nguyen\inst{11}, M.~Palutan\inst{2},
E.~Pasqualucci\inst{9}, A.~Passeri\inst{11},
V.~Patera\inst{2,8}, F.~Perfetto\inst{6},
L.~Pontecorvo\inst{9}, M.~Primavera\inst{4},
P.~Santangelo\inst{2}, E.~Santovetti\inst{10},
G.~Saracino\inst{6}, B.~Sciascia\inst{2},
A.~Sciubba\inst{2,8}, F.~Scuri\inst{7},
I.~Sfiligoi\inst{2}, T.~Spadaro\inst{2},
M.~Testa\inst{9}, L.~Tortora\inst{11},
P.~Valente\inst{9}, B.~Valeriani\inst{3},
G.~Venanzoni\inst{2}, S.~Veneziano\inst{9},
A.~Ventura\inst{4}, R.Versaci\inst{2},
G.~Xu\inst{2,1}
}

\institute{Permanent address: Institute of High Energy 
Physics of Academica Sinica, Beijing, China.\and
Laboratori Nazionali di Frascati dell'INFN, 
Frascati, Italy.\and
Institut f\"ur Experimentelle Kernphysik, 
Universit\"at Karlsruhe, Germany.\and
Dipartimento di Fisica dell'Universit\`a e Sezione INFN,
Lecce, Italy.\and
Permanent address: Institute for Theoretical 
and Experimental Physics, Moscow, Russia.\and
Dipartimento di Scienze Fisiche dell'Universit\`a 
``Federico II'' e Sezione INFN, Napoli, Italy\and
Dipartimento di Fisica dell'Universit\`a e Sezione
INFN, Pisa, Italy.\and
Dipartimento di Energetica dell'Universit\`a 
``La Sapienza'', Roma, Italy.\and
Dipartimento di Fisica dell'Universit\`a ``La Sapienza''
e Sezione INFN, Roma, Italy.\and
Dipartimento di Fisica dell'Universit\`a ``Tor Vergata''
e Sezione INFN, Roma, Italy.\and
Dipartimento di Fisica dell'Universit\`a ``Roma Tre''
e Sezione INFN, Roma, Italy.\and
Physics Department, State University of New 
York at Stony Brook, USA.\and
Physics Department, University of Virginia, USA.
}
\date{Received: date / Revised version: date}
\abstract{
We have studied the Dalitz plot of the $\ep\el\to\pio\pio\gamma$ events
collected at $\sqrt{s} \simeq M_{\phi}$ with the KLOE detector.
In the dipion invariant mass (\mpp) region below 700 MeV, the process 
under study is dominated by the non-resonant process $\ep\el\to\omega\pio$
with $\omega\to\pio\gamma$ whereas,
for higher $M_{\pi\pi}$ values, the radiative $\phi$ decay to the \fo(980)
is the dominant mechanism. Different theoretical models are used to 
fit the Dalitz plot, taking also into account a possible contribution of
the $\sigma(600)$. For each model, we extract the $\fo(980)$ mass and 
its coupling to $\pi\pi$, $\kkbar$ and to the $\phi$.
\PACS{
      {PACS-key}{discribing text of that key}   \and
      {PACS-key}{discribing text of that key}
     } % end of PACS codes
} %end of abstract
\begin{document}
%%%%%%%%%%%%%%%%%%%%%%%%%%%%%%%%%%%%%%%%%%%%%%%%%%%%%%%%%%%%%%%%%%%%%%%%%%%%%%%

\hugehead

\maketitle

%==============================================================================
\section{Introduction}
%==============================================================================

Interest in light scalar mesons remains intense in hadron spectroscopy 
due to a lack of elucidation on their nature.
There is a possibility that some of them are, in fact, exotic particles. 
There are several models to describe their 
structure, such as ordinary \qq\ mesons, \qqqq\ states or \kkbar\ molecules 
\cite{Tornqvist,Jaffe,Molecule}. 
Operating at the \ep\el\ Frascati $\phi$-factory \dafne\ \cite{DAFNE},
the KLOE experiment \cite{KLOE} is ideally suited for the study of these 
particles, since the radiative decays of the $\phi$ into two pseudoscalar 
mesons is dominated by a scalar meson ($S$) exchange in the intermediate 
state ($\phi\to S\gamma\to\pi\pi\gamma/\eta\pi\gamma/\kkbar\gamma$).
For the $\pio\pio\gamma$ final state, the possible scalar contributions 
are from the well established \fo (980) and from the more 
controversial $\sigma(600)$, purportedly observed by the E791 and BES 
collaborations \cite{s600_E791,s600_BES}.
The non-resonant \wpn\ reaction also contributes to the same final state.

The $\pio\pio\gamma$ final state had been studied at KLOE using 
16 pb$^{-1}$ of 2000 data \cite{PLBf0n2002}.
The resulting ratio between the $\fo KK$ and $\fo\pi\pi$ couplings, 
together with the large value of the \br, favoured the \qqqq\ composition 
of the \fo (980), while the shape of the \pio\pio\ invariant mass 
suggested a possible contribution also from $\sigma(600)$.
In the present paper, the analysis is repeated with a statistics  
about thirty times larger, thus allowing us to study this reaction in much
greater detail.
A common set of cuts and algorithms for the resonant and non-resonant 
processes has been developed so that,
by fitting the Dalitz plot, the differential cross sections of the 
two components are extracted.
A detailed technical description of this analysis is in 
Refs.~\cite{KNana,KNfit}.

%==============================================================================
\section{Experimental setup}
%==============================================================================

\begin{figure}[!t]
  \resizebox{\columnwidth}{!}{\includegraphics{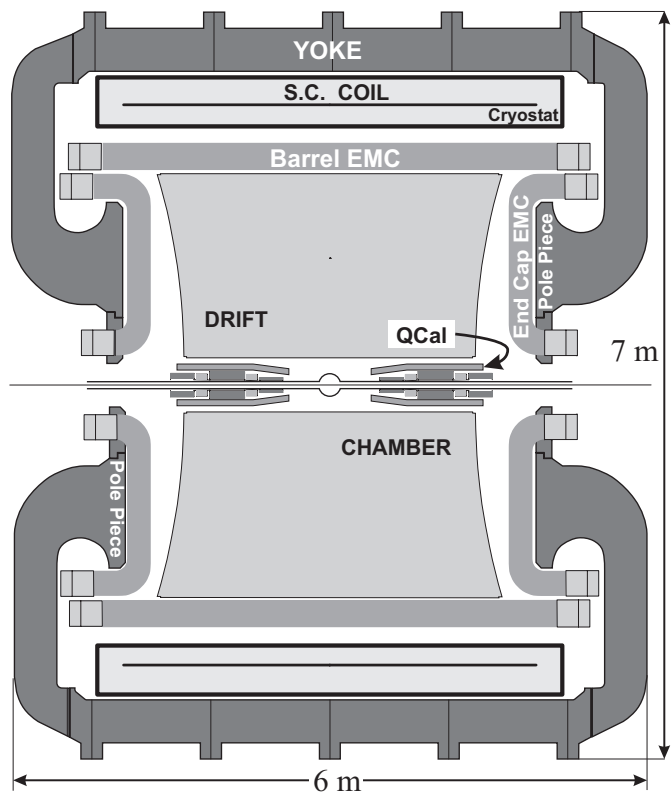}}
  \caption{Vertical cross section of the KLOE detector.}
  \label{Fig:KLOE}
\end{figure}

Data were collected with the KLOE detector at \dafne, the Frascati 
$\ep\el$ $\phi$-factory, which operates at a center of mass energy 
$\roots = M_\phi\sim 1020$ MeV. The beams collide with a crossing 
angle of $(\pi - 25)$ mrad, producing $\phi$ mesons with a small
momentum ($p_\phi \sim 13$ MeV/c) in the horizontal plane.
The KLOE detector (see Fig.~\ref{Fig:KLOE}) is inserted in a 0.52 T 
magnetic field. It consists of a 2 m radius drift chamber (DC) 
\cite{DCH}, with full stereo geometry using helium based gas mixture, 
surrounded by a fine sampling lead/scintillating fibers electromagnetic 
calorimeter (EMC) \cite{EMC}, divided into a barrel and two endcaps,
with a hermetic coverage (98\% of the solid angle) and a very high 
efficiency for low energy photons. Since the channel 
$\ep\el\to\pio\pio\gamma$ under study is 
fully neutral, its analysis is based mainly on the EMC performance. 
The arrival times of particles and the positions 
in three dimensions of the energy deposits are obtained from the 
signals collected at the two ends of the calorimeter modules, with 
a granularity of $\sim ( 4.4 \times 4.4)$~cm$^2$, for a total of
2440 cells arranged in five layers.
Cells close in time and space are grouped into a calorimeter cluster. 
The probability of a photon to fragment in more than a cluster
(splitting) is reduced by employing a special recovery algorithm.
The cluster energy $E$ is the sum of the cell energies, while the cluster 
time $T$ and its position $\vec{R}$ are energy weighted averages. 
Photon energy and 
time resolutions are $\sigma_E/E = 5.7\%/\sqrt{E\ {\rm(GeV)}}$ and  
$\sigma_T = 57\ {\rm ps}/\sqrt{E\ {\rm(GeV)}} \oplus 100\ {\rm ps}$, 
respectively. 
The KLOE trigger \cite{TRG} is based on the detection of two energy deposits 
(called sectors) with $E>50$ MeV for barrel and $E>150$ MeV for endcaps. 
Events with only two fired trigger sectors in the same endcap are rejected,
being this topology dominated by machine background. 
Recognition and rejection of cosmic-ray events is also performed at the 
trigger level, selecting events with two energy deposits above a 30 MeV 
threshold in the outermost calorimeter layer.
Moreover, to reject residual cosmic rays and machine background events, 
an offline software filter uses calorimeter and DC information before 
track reconstruction~\cite{NIMOffline}.

The machine parameters (center of mass energy \roots, $\phi$ 
momentum and beams interaction point) are measured on-line from the
analysis of Bhabha scattering events in the barrel.
The average value of the center-of-mass energy is 
evaluated with a precision of 30 keV per run, each corresponding to 
$\sim 100\ \mbox{nb}^{-1}$ of integrated luminosity.   
To calibrate the absolute beam energy scale we fit the $\phi$ 
line shape of $\phi\to\eta\gamma$ events \cite{KNana}. Comparing the
value obtained for $M_\phi$ with the precise measurement of the CMD-2 
experiment obtained with the depolarization method \cite{MphiCMD2}, a
shift of $+150$~keV is found and is corrected in our analysis accordingly.

Prompt photons are identified as neutral particles with $\beta=1$
originated at the interaction point, by requiring 
$|T-R/c|<{\rm min}(5\,\sigma_T, 2\ {\rm ns})$,
where $T$ is the photon flight time and $R$ the corresponding path length.
The photon detection efficiency is $\sim 90\%$ for $E_\gamma$=20~MeV, and 
reaches 100\% above 70~MeV.
The sample selected by the timing requirement has less than $0.6\%$
contamination per event due to accidental clusters from machine background.

%==============================================================================
\section{Event selection}
%==============================================================================

All the available statistics collected in 2001--2002 data-taking 
periods, corresponding to 450 pb$^{-1}$, has been analyzed by 
grouping all runs in center of mass energy bins of 100 keV.
This was done to take into account the $\sim 2$ MeV spread in the center 
of mass energy present in the data set.
For this analysis, only those runs belonging to the bin with the highest 
statistics have been used for fitting the Dalitz plot.
This sample corresponds to 145 pb$^{-1}$ collected at 
$\roots = ( 1019.7 \div 1019.8 )$~MeV.

The response of the detector to the decays of interest was studied by 
using the KLOE Monte Carlo (MC) simulation program \cite{NIMOffline}. 
The MC 
takes into account changes in the machine operation and background conditions,
in order to reproduce real data on a run-by-run basis. For the present 
analysis, an MC sample for both signal and backgrounds is produced. The 
corresponding integrated luminosity is five times that of the collected 
data, except for $\ep\el\to\gamma\gamma$ events that are produced at
a 1:1 rate. For the simulation of signal events, the \mpp\ spectrum for 
the $\phi\to S\gamma\to\pi\pi\gamma$ ($S\gamma$) process is produced according
to the shape obtained from 2000 data, while the \wpn\ ($\wpi$) generator is 
based on the Vector Meson Dominance (VMD) description of the three body 
decay according to Ref.~\cite{SwpnSND}.

The data analysis consists of four steps: 
\begin{enumerate}
\item an acceptance selection of five prompt photons with 
  $E_\gamma \ge 7$ MeV and a polar angle satisfying the requirement
  $|\cos\theta_\gamma| < 0.92$;
\item a kinematic fit (Fit1) imposing total 4-momentum conservation;
\item a pairing procedure of photons to \pio's,
  where the photon combination minimizing a pseudo-$\chi^2$ built using 
  the invariant mass of the two $\gamma\gamma$ pairs, $\chi^2_{\rm pair}$, 
  is selected as the good one;
\item a second kinematic fit (Fit2), where the constraints on the 
  \pio\ masses are also imposed. 
  The selected events must then satisfy the requirements 
  $\chi^2_{\rm Fit2}/{\rm Ndf}\le 5$ and $\Delta M_{\gamma\gamma}= 
  |M_{\gamma\gamma}-M_{\pi}| \le 5\,\sigma_{\gamma\gamma}$, where 
  $M_{\gamma\gamma}$ and $\sigma_{\gamma\gamma}$ are evaluated using 
  the photon momenta from Fit1. 
\end{enumerate}

A further cut is applied to reject the background from the
$\ep\el\to\gamma\gamma$ process, which has a 
cross section much larger than the signal and where the three additional 
prompt photons could be generated either by radiation, by cluster
splitting or by accidental coincidence with machine background clusters.
Such process is hugely reduced without losing efficiency 
for the signal, by rejecting events where the energy sum of the two most 
energetic clusters in the event is greater than 900 MeV.

\begin{figure}[!t]
  \resizebox{\columnwidth}{!}{\includegraphics{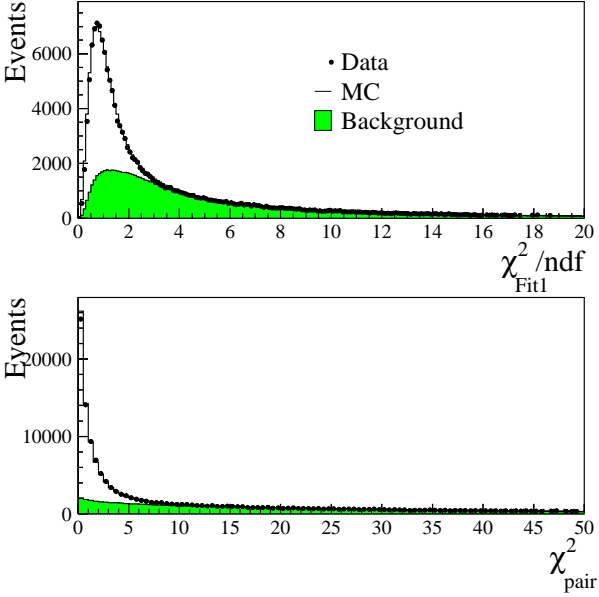}}
  \caption{Data-MC comparison after acceptance selection: normalized 
    $\chi^2$ from the first kinematic fit (top) and the minimum value
    of the pseudo-$\chi^2$ used to pair photons (bottom).}
  \label{Fig:Chi2Fit1}
\end{figure}

\begin{figure}[!t]
  \resizebox{\columnwidth}{!}{\includegraphics{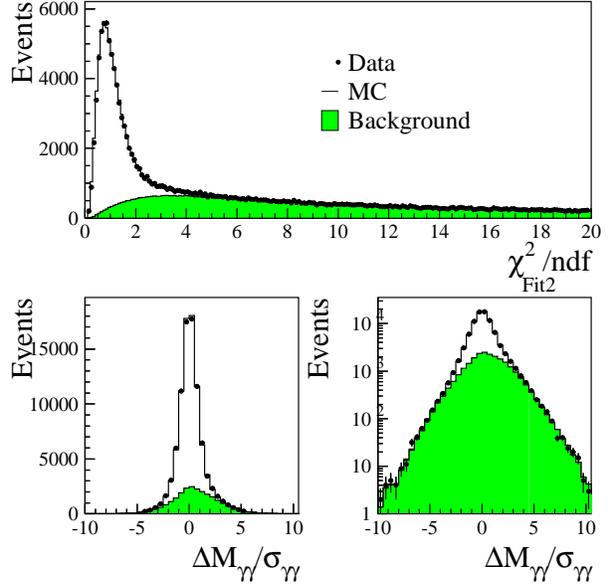}}
  \caption{Top: normalized $\chi^2$ of the second kinematic fit after 
    acceptance cuts. Bottom: normalized $\Delta M_{\gamma\gamma}$ of the 
    selected $\gamma\gamma$ pairs after $\chi^2_{\rm Fit2}$ cut in linear 
    (left) and logarithmic (right) scale.}
  \label{Fig:Selection}
\end{figure}

The overall analysis efficiency for the identification of the signal is 
evaluated by applying the whole analysis chain to the $S\gamma$ and \wpi\ 
MC events: $\varepsilon_{S\gamma} = (50.3 \pm 0.1)\%$, 
$\varepsilon_{\omega\pi} = (53.12\pm 0.05)\%$. The small difference is 
due to the characteristic energy and angular distributions of
photons in the two kinds of events.
After acceptance selection we start with a sample of 243,904 events of 
which 86,449 survive the complete analysis chain. 
As shown in Figs.~\ref{Fig:Chi2Fit1}, \ref{Fig:Selection} excellent 
data-MC agreement is found, both after acceptance selection and after 
applying the complete analysis chain.

\begin{figure}[!t]
  \resizebox{\columnwidth}{!}{\includegraphics{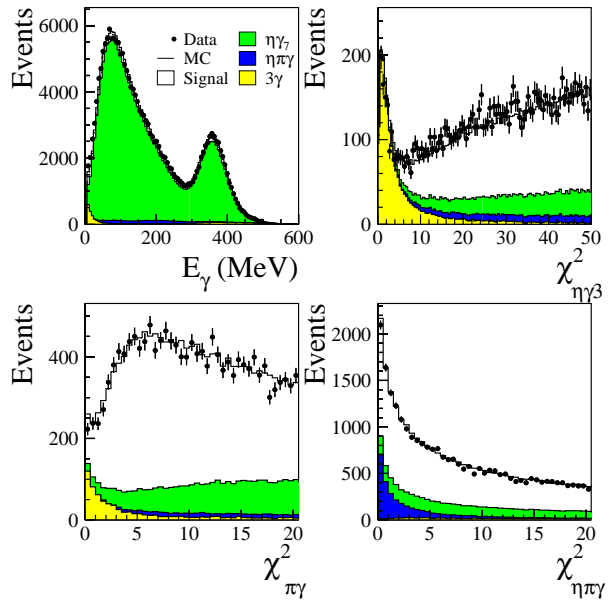}}
  \caption{Distributions used to check background contamination.
    Top-left: photon energy distribution for events with 
    $4<\chi^2_{\rm Fit2}/{\rm ndf}<20$. Minimum value of the 
    $\chi^2$ used to pair photons in the $\eta\gamma_3$ 
    (top-right), $\pi\gamma$ (bottom-left) and $\eta\pi\gamma$ 
    (bottom-right) hypotheses. The $3\gamma$ distributions include
    $\eta\gamma_3$, $\pi\gamma$ and $\gamma\gamma$ processes.}
  \label{Fig:Background}
\end{figure}

The background channels for the $\pio\pio\gamma$ final state are
listed in Tab.~\ref{Tab:SB_ppg} together with the analysis efficiency 
and the corresponding signal to background ratio,
before and after the application of the analysis cuts, evaluated 
using branching ratios (\br s) from Ref.~\cite{PDG03} or KLOE 
measurements whenever available \cite{PLBa0,eta2000}.
The $\phi\to\eta\pi^0\gamma$ channel ($\eta\pi\gamma$) is the only one that
has the same five photon final state as the signal.
The $\phi\to\eta\gamma\to\pio\pio\pio\gamma$ ($\eta\gamma_7$)
mimics the signal when there are lost or merged photons. The three
photon final states ($\phi\to\eta\gamma$ [$\eta\gamma_3$],
$\phi\to\pio\gamma$ [$\pi\gamma$] and $\ep\el\to\gamma\gamma(\gamma)$
[$\gamma\gamma$]) produce five clusters due to splitting or accidental
coincidence with clusters produced by machine background.
We have used background-enriched distributions to check
the absolute yields and the Monte Carlo shapes.
Each data distribution ($\rm H_{data}$) has been fit
with two MC components: the background under consideration ($\rm H_{bckg}$) 
and all the other contributions, including the 
$\pio\pio\gamma$ signal ($\rm H_{others}$):
$\rm H_{data} = \alpha_1\, H_{bckg} + \alpha_2\, H_{others}$.
These distributions, used to fit the main background components 
after applying the $\alpha_1$ scale factors, are shown in 
Fig.~\ref{Fig:Background}. 
The dominant contribution $\eta\gamma_7$ is verified by
studying the events in the region $4<\chi^2_{\rm Fit2}/{\rm ndf}<20$.
For all other background sources, we keep the standard analysis 
cuts and build a specific $\chi^2$, minimizing the difference between
the reconstructed and the true mass of the intermediate particles 
($\eta$ and \pio) in the corresponding hypothesis.
The values of $\alpha_1$ obtained for all the backgrounds are listed in 
Tab.~\ref{Tab:SB_ppg} together with the uncertainties on the \br s.
We do not apply these scale factors, but we use them in the evaluation
of the fit systematics as discussed in Par.~\ref{Par:Syst}.
Note that for the dominant contribution ($\eta\gamma_7$) a scale
factor statistically consistent with the expected rate is found.

\begin{table*}[!t]
  \caption{Analysis efficiency for background events. The signal 
    over background ratio before (S/B) and after (S/B$_{\rm ana}$) 
    the application of the analysis cuts is also shown. In the last 
    two columns the error on the expected rate due to the uncertainties
    on the used \br s and the scale factor obtained from 
    the fit to the background enriched distributions are reported.}
  \label{Tab:SB_ppg}
  \begin{center}    
    \renewcommand{\tabcolsep}{5mm}
    \renewcommand{\arraystretch}{0.9}    
    \begin{tabular}{|l|ccc|cc|}  \hline
      Process   &   $\varepsilon_{\rm ana}$   & S/B   &  
      S/B$_{\rm ana}$ & $\delta\br/\br$ & $\alpha_1$ \\ \hline
      $\phi\to\eta\pio\gamma\to\gamma\gamma\pio\gamma$  &
      $(23.2\phantom{5} \pm 0.1\phantom{5}) \%$ &  8.5 &  19 & 
      \phantom{1}9.5\%  &   $0.86\pm 0.02$    \\
      $\phi\to\eta\gamma\to\pio\pio\pio\gamma$   &
      $(8.51\pm 0.02) \cdot 10^{-3}$            &  0.06 &  4 & 
      \phantom{1}3.0\%  &   $1.064 \pm 0.002$ \\
      $\phi\to\eta\gamma\to\gamma\gamma\gamma$   &
      $(8.15 \pm 0.05)\cdot 10^{-4}$            &  0.06 &  30 & 
      \phantom{1}3.5\%  &   $0.86\pm 0.02$    \\
      $\phi\to\pio\gamma$                        &
      $(3.07 \pm 0.06)\cdot 10^{-4}$            &  0.2 & 350 & 
      10.0\%            &  $2.35\pm 0.02$     \\
      $\ep\el\to\gamma\gamma$                    &
      $(0.19 \pm 0.01)\cdot 10^{-5}$            & 0.002 & 400 &  
      ---               &  $1.85\pm 0.03$     \\ \hline
    \end{tabular}
  \end{center}
\end{table*}

\begin{figure}[!t]
    \resizebox{\columnwidth}{!}{\includegraphics{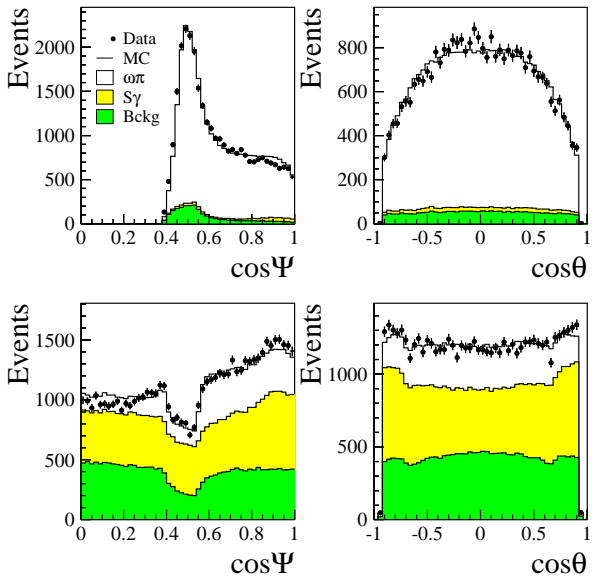}}
  \caption{Data-MC comparison for \wpi\ (top) and $S\gamma$ (bottom) 
    events selected by looking at the mass of the intermediate state 
    and assuming the interference term to be negligible. 
    Left panels: $\cos\psi$ distributions.
    Right panels: $\cos\theta$ for the primary photon.}
  \label{Fig:ExclAngles}
\end{figure}

In order to check the relative contribution of the two signal processes, 
their angular distributions are studied.
Both the photon polar angle ($\theta$) and the minimum angle 
between the photon and the \pio 's in the \pio\pio\ rest frame ($\psi$)
are expected to show a different behaviour due to the spin 
of the intermediate particles involved. 
To first order the interference between the two processes 
can be neglected and they can be separated on event by event basis 
by looking at the mass of the intermediate state. The $\omega$ mass is 
reconstructed by selecting the best match of the two $\pio\gamma$ 
combinations. After background subtraction, events with
$|M_{\pi\gamma}-M_{\omega}|<3\,\sigma_{M_\omega}$%
\footnote{$\sigma_{M_\omega} = 9.5$ MeV is the convolution between the 
experimental resolution and the $\omega$ width.}
are classified as \wpi, while all the rest is called $S\gamma$.
In Fig.~\ref{Fig:ExclAngles} the data-MC comparison for the $\cos\theta$ 
and $\cos\psi$ angular distributions are shown for both $S\gamma$ and 
\wpi\ processes. 
The simple superposition of the $S\gamma$ and $\wpi$ MC shapes fits
rather well the data, suggesting that the contribution of the interference 
term is small.

\begin{figure}[!t]
  \begin{tabular}{cc}
    \hspace{-1cm}
    \resizebox{0.55\columnwidth}{!}{\includegraphics{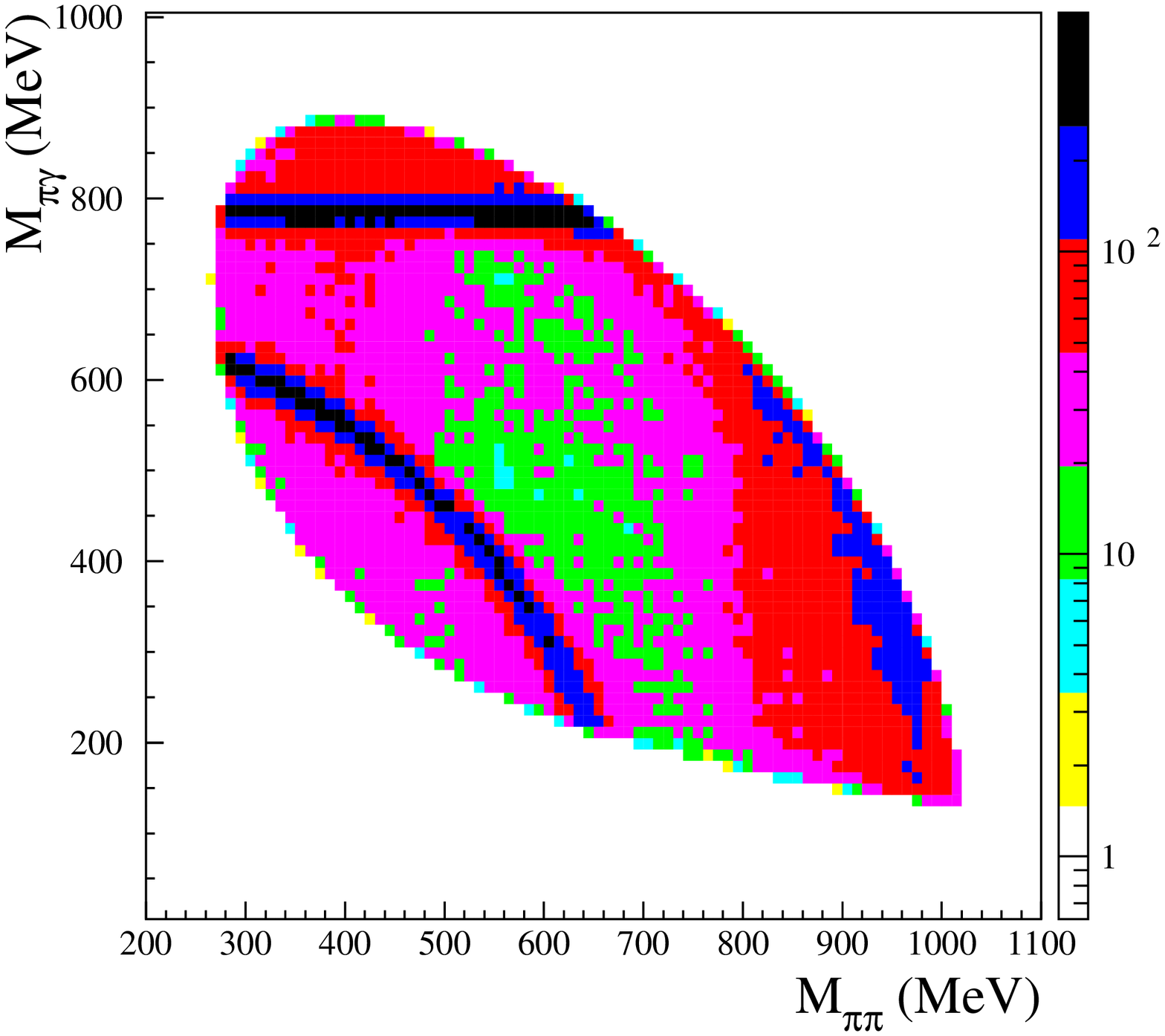}} &
    \hspace{-0.5cm}
    \resizebox{0.55\columnwidth}{!}{\includegraphics{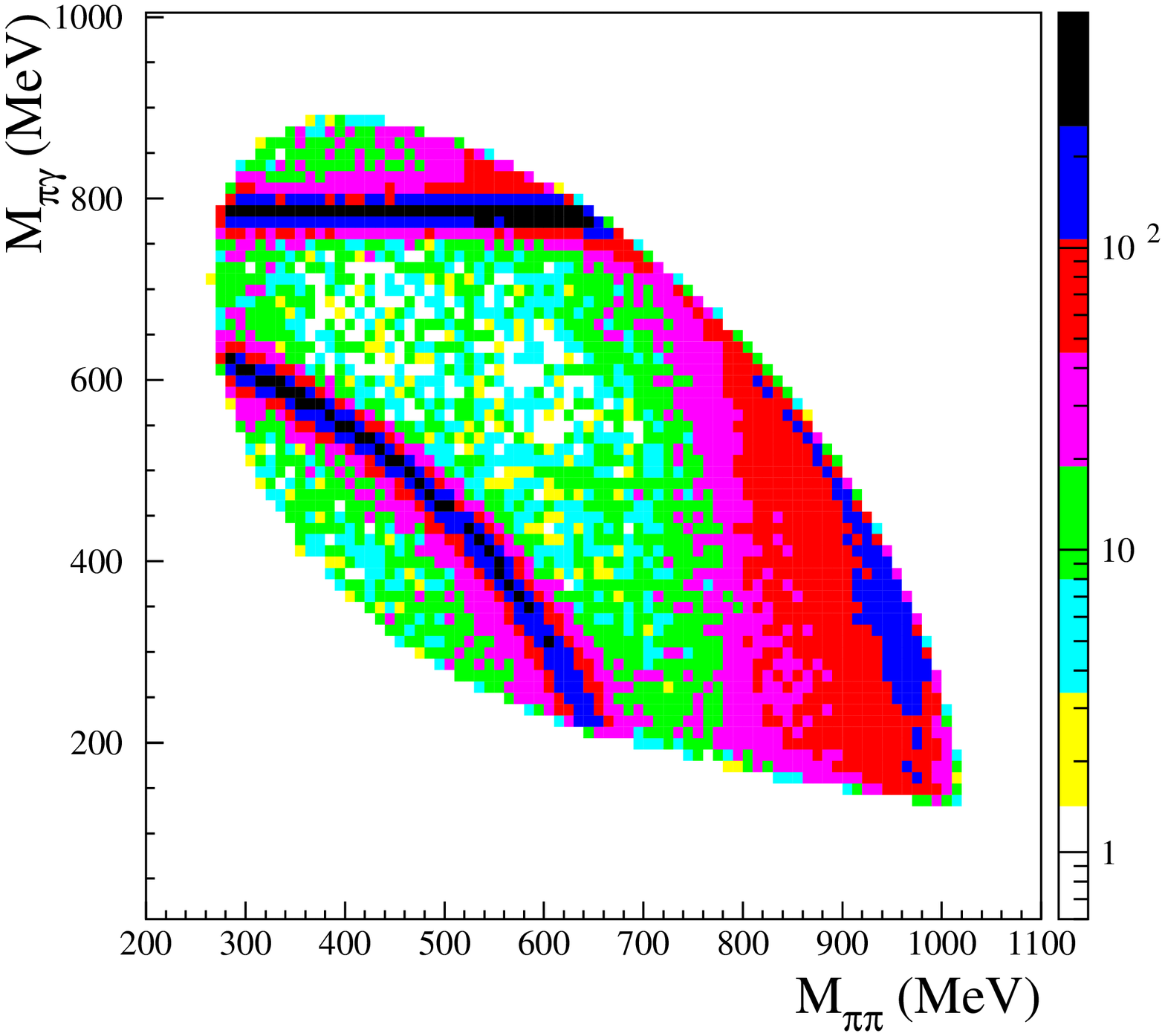}} \\
  \end{tabular}
  \caption{Dalitz plot in logarithmic scale before (left) and after 
    (right) background subtraction.
    The two bands in the region $\mpp < 700$ MeV are due to 
    $\ep\el\to\omega\pio$ events.}
  \label{Fig:DalitzLog}
\end{figure}

\begin{figure}[!t]
  \resizebox{\columnwidth}{!}{\includegraphics{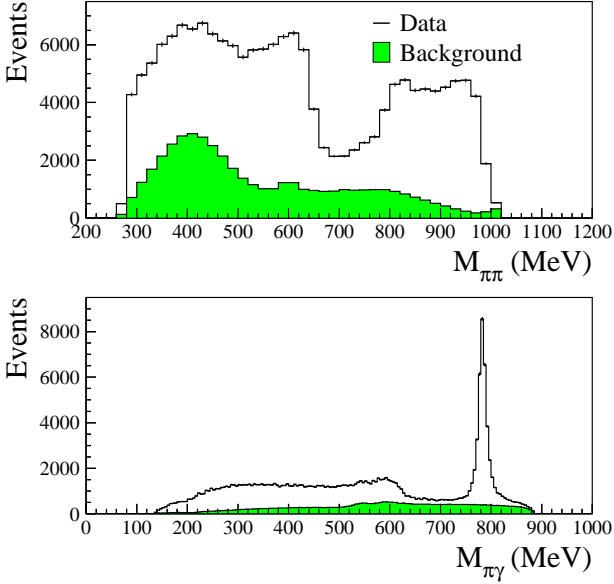}}
  \caption{Dalitz plot projections in \mpp\ (top) and \mpg\ (bottom) 
    variables before background subtraction.}
  \label{Fig:DalitzProj}
\end{figure}

The two kinematic variables chosen for the construction of the Dalitz 
plot are the invariant masses of the two \pio's, \mpp, and of the two 
possible $\pio\gamma$ combinations, \mpg. We have therefore two entries 
per event. The binning choice, driven by the mass resolutions obtained 
by Monte Carlo for the signal, is 10 MeV for \mpp\ and 12.5 MeV for \mpg.
The data density is shown in Fig.~\ref{Fig:DalitzLog} before and after 
background subtraction. 
The two projections are shown in Fig.~\ref{Fig:DalitzProj}.
After background subtraction the number of events in the Dalitz plot 
is $128,529\pm 659$. 

\begin{figure}[!t]
    \resizebox{\columnwidth}{!}{\includegraphics{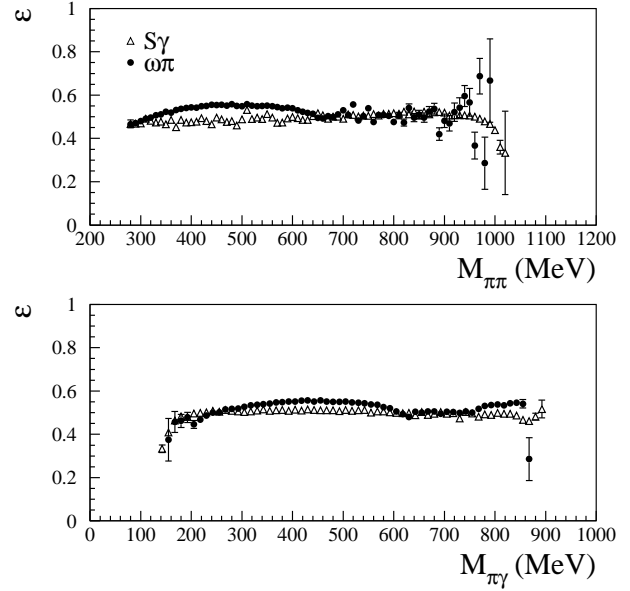}}
  \caption{Analysis efficiency for $S\gamma$ (empty triangles) and \wpi\ 
    (black dots) MC events as a function of \mpp\ (top) and \mpg\ (bottom).}
  \label{Fig:Effi}
\end{figure}

The analysis efficiency as a function of \mpp\ and \mpg\ is evaluated 
by Monte Carlo, with corrections based on data control samples.
In Fig.~\ref{Fig:Effi} the dependence of the selection efficiency on
\mpp\ and \mpg\ is shown for the $S\gamma$ and \wpi\ final states. 
Both processes exhibit a rather flat dependence on the Dalitz plot variables. 
The different shape is related to their angular distribution and to their 
different probability of producing photons from initial state radiation (ISR).
The main source of data-MC differences is due to the photon detection 
efficiency, which is measured as a function of $E_{\gamma}$ with 
$\phi \to \pi^+\pi^-\pi^0$ control samples and applied, as a correction, 
to the Monte Carlo. 
The MC trigger, cosmic ray veto and event classification filter efficiencies 
are checked using prescaled data samples. The overall correction factor, 
$R_{\rm sel}= 1.022 \pm 0.004$, is applied to the MC analysis efficiency.

%==============================================================================
\section{Fit to the Dalitz plot}
%==============================================================================

%------------------------------------------------------------------------------
\subsection{Fitting method}
%------------------------------------------------------------------------------
\label{Sec:FitMethod}

\begin{figure}[!t]
  \resizebox{\columnwidth}{!}{\includegraphics{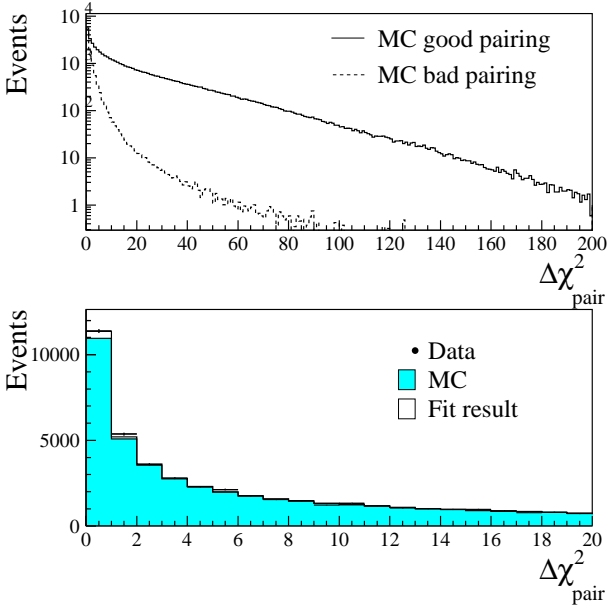}}
  \caption{Top: Monte Carlo $\Delta\chi^2_{\rm pair}$ shapes for events 
    with correct and wrong photon pairing. These two templates are used
    to fit the $\Delta\chi^2_{\rm pair}$  data distribution (bottom).
    A better data-MC agreement is obtained after the fit.}
  \label{Fig:Chi2SelFit}
\end{figure}

The double differential $\pio\pio\gamma$ cross section is written as 
the sum of the scalar term, the VMD contribution and of their relative 
interference as follows:
\begin{eqnarray}
 && \frac{d\sigma}{d\mpp dM_{\pi\gamma}} = \label{Eq:Xsec} \\ \nonumber
 && \ \ \ \frac{d\sigma^{S}}{d\mpp dM_{\pi\gamma}} +
 \frac{d\sigma^{V}}{d\mpp dM_{\pi\gamma}} \pm
 \frac{d\sigma^{I}}{d\mpp dM_{\pi\gamma}}.
\end{eqnarray}
The suffixes $S$ and $V$ stand for $S\gamma$ and VMD terms while $I$ 
represents the interference. To fit the Dalitz plot data density, the 
theory is folded with the reconstruction efficiencies of the 
two processes and with the probability for an event to migrate from 
a Dalitz plot bin to another one. The expected number of events for a 
given reconstructed \mpp, \mpg\ bin $i$, $N^{\rm exp}_{i}$, is then 
computed from the total integrated luminosity, \lint, as follows:
\begin{eqnarray}
 && N^{\rm exp}_{i} = \lint \times \\
 && \ \ \ \ \sum_{j} \, ( f^{V}_{j} \ A^{V}_{i,j} \ \varepsilon^{V}_{j} +
 f^{S}_{j} \ A^{S}_{i,j} \ \varepsilon^{S}_{j} +
 f^{I}_{j} \ A^{V}_{i,j} \ \varepsilon^{V}_{j}) \, . \nonumber
\end{eqnarray}
where $f_{j}$ is the integration of the differential cross section 
evaluated in the $j^{\rm th}$ bin, including the effect of the ISR. 
For each bin, $\varepsilon_{j}$ is the analysis efficiency while $A_{i,j}$ 
is the smearing matrix, representing the probability for the signal 
event to migrate from the $j^{\rm th}$ to the $i^{\rm th}$ bin, either 
due to resolution or to a wrong reconstruction of the event. 
This matrix has been evaluated by Monte Carlo: about 85\% of the events 
are on the diagonal or close to it, within $\pm 1$ bin. The remaining 
15\% is mostly due to events where photons are incorrectly paired to \pio's. 
A dedicated data-MC comparison has been performed to calibrate the 
fraction of good/bad pairing.
The difference between the two lowest values of pseudo-$\chi^2$ used 
to pair photons, $\Delta\chi^2_{\rm pair}$, is fit with a superposition
of the two templates 
obtained for MC events with good and bad photon pairing 
(Fig.~\ref{Fig:Chi2SelFit}). A data-MC difference of $1.08 \pm 0.02$ 
is found and is taken into account in evaluating the systematic error.

%------------------------------------------------------------------------------
\subsection{Theoretical models}
%------------------------------------------------------------------------------

A more explicit formulation of the differential cross section 
(\ref{Eq:Xsec}) is reported in Appendix \ref{App:Xsec}.
Concerning the scalar term, we use two different approaches for the 
description of the amplitude: the Kaon Loop model (KL)
\cite{KaonLoop,KaonLoopSigma,KaonLoopNew}, where the scalar is 
coupled to the $\phi$ through a charged kaon loop, and 
another formulation called No Structure (NS) \cite{NoStructure},
where
the $\phi S\gamma$ coupling is point-like and the scalar is described 
as a Breit-Wigner with a mass dependent width added to a %polynomial 
complex function, to allow an appropriate behaviour of the resulting 
shape at low masses.

\subsubsection{Kaon Loop model}

As scalar term of the Kaon Loop model we use the one described in 
Ref.~\cite{KaonLoopNew}, where the amplitude of the signal
\begin{equation}
  M_{S\gamma} = e^{i \delta_B} g(M_{\pi\pi}) \, \left( \sum_{S,S'} 
    g_{SK^+K^-} \, G^{-1}_{SS'} \, g_{S'\pi^0\pi^0} \right) ,
\end{equation} 
implies the mixing of two scalar states, namely the \fo(980) 
and the $\sigma(600)$, represented by the $G_{SS'}$ matrix.
The $g(M_{\pi\pi})$ function describes the kaon loop and $g_{S\pi\pi/KK}$ 
are the couplings of the scalars to the $\pi\pi/\kkbar$ mesons.
Differently from the past, where the phase $\delta_B$ took into account 
only the elastic scattering due to $\pi\pi$, in this new
formulation the scattering due to \kkbar\ is also considered.

Using this theoretical framework, our \pio\pio\ mass spectrum obtained 
from 2000 data \cite{PLBf0n2002}, has been fit by the authors of 
Ref.~\cite{KaonLoopNew} together with $\pi\pi$ scattering data 
\cite{ScattData1,ScattData2,ScattData3,ScattData4,ScattData5}, providing
ten sets of parameters which are able to describe both data samples.
For all of them, the $\sigma(600)$ 
coupling to $K\overline{K}$ is small with respect to the one of $\pi\pi$ 
and its mass lies in the 500--700 MeV range. This means a very 
broad meson width, between 240 and 490 MeV. 
All the odd (even) variants have a positive (negative) 
$g_{\sigma\pi^{+}\pi^{-}}/g_{f_{0}K^{+}K^{-}}$ ratio.

When fitting the Dalitz plot of our new data set, we can not leave all 
$\fo(980)$ and $\sigma(600)$ parameters free, as the description of 
the elastic background
and the couplings and masses of the $\sigma(600)$ meson are closely related. 
We therefore proceed by using as free parameters only the VMD ones
and the mass and the coupling to \pip\pim\ and \kkbar\ of the \fo(980),
using the isospin relations 
$g_{f_{0}\pi^0\pi^0} = 1/\sqrt{2} g_{f_{0}\pi^+\pi^-}
= 1/\sqrt{3}g_{f_{0}\pi\pi}$ and $g_{f_{0}KK} = g_{f_{0}K^+K^-} = 
g_{f_{0}K^0\overline{K}^0}$.
For the $\sigma(600)$ and the elastic $\pi\pi$ 
and \kkbar\ scattering we use the values of Ref.~\cite{KaonLoopNew} by 
repeating the fit for all the ten available sets of parameters. 
We obtain $P(\chi^2)$ ranging from $10^{-5}$ to 0.145; only the six 
results with $P(\chi^2)>1\%$ are kept for the present discussion.%
\footnote{The accepted results are the variants 1, 2, 3, 8, 9, 10 of
Ref.~\cite{KaonLoopNew}.}

Furthermore, the case of a single scalar contribution, the \fo(980), in the 
Kaon Loop description has also been tried by using the old KL 
parametrization \cite{KaonLoopSigma}.
The resulting $P(\chi^2)$ of the fit is not acceptable, showing the need 
of introducing the $\sigma(600)$.

\subsubsection{No Structure model}

In the NS description, the amplitude of the scalar term is proportional 
to a scalar form factor, $F_{0}^{\rm scal}$, which is derived by 
assuming a direct coupling of the $\phi$ to the $\fo(980)$, 
$g_{\phi f_0 \gamma}$, and a subsequent coupling of the $\fo(980)$ to 
the $\pi\pi$ pair, $g_{f_0\pi\pi}$. In the same form factor the possible 
non-resonant continuum background is also added as a series expansion in 
$M_{\pi\pi}$ as follows:
\begin{eqnarray}
  F_{0}^{\rm scal} & = & \frac{g_{f_0 \pi^0\pi^0} \, g_{\phi f_0 \gamma}}
  {M_{\pi\pi}^2-M_{f_0}^2 +i\, M_{\pi\pi}\Gamma_{f_0}(M_{\pi\pi})} + \\
  & & \frac{a_0}{M_{\phi}^2} e^{i b_0 \frac{v_{\pi}(M_{\pi\pi})}{M_{\phi}}}
  + \frac{a_1}{M_{\phi}^4}
  e^{i b_1 \frac{v_{\pi}(M_{\pi\pi})}{M_{\phi}}}(M_{\pi\pi}^2-M_{f_0}^2), 
  \nonumber
  \label{Eq:FscalNS}
\end{eqnarray}
where, in the most general case, the background parameters are complex 
numbers and  $v_{\pi}(M_{\pi\pi})$ is proportional to pion momentum
in the scalar rest frame, $v_{\pi} = \sqrt{M_{\pi\pi}^2/4-M_{\pi}^2}$.
The propagator for the \fo(980) resonance is described by a simple
Breit-Wigner shape corrected by the Flatt\`e condition on the 
$\pi\pi$ and $KK$ thresholds, i.e.:
\begin{eqnarray}
  \Gamma_{f_0}(M_{\pi\pi}) & = & g_{f_0\pi\pi}^2 
  \frac{v_{\pi}(M_{\pi\pi})}{8\pi M_{\pi\pi}^2} + \\ \nonumber
  & & g_{f_0KK}^2 
  \frac{v_{K^\pm}(M_{\pi\pi})+ v_{K^0}(M_{\pi\pi})}{8\pi M_{\pi\pi}^2},
\end{eqnarray}
where $v_{\pi,K,K^0}$ are complex numbers with an 
analytical continuation under threshold and the coupling to $\pi\pi$
and \kkbar\ have the same meaning as in the KL description.
In this model the fit parameters are $M\subfo$, $g_{f_{0}\pi^+\pi^-}$,
$g_{f_{0}KK}$, $g_{\phi f_{0}\gamma}$ and the background parameters
$a_0$, $a_1$ and $b_1$. The $b_0$ phase is fully determined as a 
function of the other parameters. 

%------------------------------------------------------------------------------
\subsection{Fit Systematics}
%------------------------------------------------------------------------------
\label{Par:Syst}

There are different sources of systematics affecting this analysis which 
can give rise to variations of the fit results. We describe here the most 
important ones. For each of them, the fit is repeated after changing 
the relevant quantity within its range of uncertainty. 
In Tabs.~\ref{Tab:SystKL}, \ref{Tab:SystNS} we show the corresponding 
percentage variation of the interesting free parameters for the KL, NS 
model respectively. 

\begin{table*}[!t]
  \caption{Fractional systematic error on fit parameters for the KL scalar 
    term. Only variations above 0.1\% are reported.}
  \label{Tab:SystKL}
  \begin{center}
    \renewcommand{\tabcolsep}{5mm}
    \renewcommand{\arraystretch}{0.9}    
    \begin{tabular}{|l|ccc|} \hline
      Parameter & $M\!\subfo$ & $g_{\!f_{0}K^{+}K^{-}}$ & 
      $g_{\!f_{0}\pi^{+}\pi^{-}}$ \\ \hline\hline
      Source & \multicolumn{3}{|c|}{Fractional systematic error}   \\ \hline
      Normalization      & ---      & $\pm 1.6\%$ & $\pm 0.7\%$     \\
      Beam energy        & ---      & $+2.4\%$    & $-1.4\%$        \\
      Photon efficiency  & ---      & $-2.1\%$    & $+1.4\%$        \\
      $\chi^2$ cut       & ---      & $-0.8\%$    & ---             \\
      Smearing matrix    & $-0.1\%$ & ---         & $+1.4\%$        \\ 
      Interference       & ---      & $-1.1\%$    & $+0.7\%$        \\
      Background         & $+0.1\%$ & $+4.0\%$    & $-3.5\%$        \\ \hline
   \end{tabular}
  \end{center}
\end{table*}

\begin{table*}[!t]
  \caption{Fractional systematic error on fit parameters for the NS scalar 
    term. Only variations above 0.1\% are reported.}
  \label{Tab:SystNS}
  \begin{center}
    \renewcommand{\arraystretch}{0.9}    
    \begin{tabular}{|l|ccccccc|} \hline
      Parameter & $M\!\subfo$ & 
      $g_{\!f_{0}K^{+}K^{-}}$ & 
      $g_{\!f_{0}\pi^{+}\pi^{-}}$  & 
      $g_{\phi f_0 \gamma}$  & 
      $a_0$ & $a_1$ &  $b_1$ \\ \hline\hline
      Source & \multicolumn{7}{|c|}{Fractional systematic error}      \\ \hline
      Normalization   & $\pm 0.1\%$ & $^{+155\%}_{-\phantom{1}50\%}$ &
      $^{+6.9\%}_{-1.5\%}$ & $^{+11.9\%}_{-\phantom{1}1.9\%}$ &
      $^{+34.4\%}_{-\phantom{3}5.5\%}$ & $^{+122\%}_{-\phantom{1}16\%}$ & 
      $^{+57.3\%}_{-16.7\%}$ \\
      Beam energy     & $-0.2\%$ & $+82.5\%$ & $+0.8\%$ & $+1.9\%$ &
      ---      & ---      & $-\phantom{2}3.3\%$ \\
      Photon efficiency & $+0.2\%$ & $-72.5\%$ & $+3.1\%$ & $+3.4\%$ &
      $+17.9\%$ & $+49.6\%$ & $+29.3\%$ \\
      $\chi^2$ cut    & $-0.2\%$ & $+57.5\%$ & $+2.3\%$ & $-2.3\%$ &
      $+\phantom{1}1.7\%$  & $+17.6\%$ & $+20.0\%$ \\
      Smearing matrix & $-0.2\%$ & $-\phantom{7}7.5\%$ & $-2.3\%$ & 
      $-3.1\%$ & $-10.0\%$ & $-26.0\%$ & $-28.0\%$ \\
      Interference    & ---      & $+82.5\%$ & $+3.1\%$ & $+5.4\%$ &
      $+18.1\%$ & $+69.5\%$ & $+46.0\%$ \\
      Background      & $-0.4\%$ & $+92.5\%$ & --- & $-2.3\%$ & 
      $-8.4\%$ & $-8.4\%$ & $+\phantom{2}4.0\%$    \\ \hline
    \end{tabular}
  \end{center}
\end{table*}

\begin{enumerate} 

\item {\em Normalization}\\
The first effect considered is the normalization scale of the fit 
estimate on the event counting. When evaluating $N^{\rm exp}_{i}$, two 
experimentally determined constants are used: the integrated luminosity 
and the value of the $\phi$ leptonic  width. The luminosity is known 
with a total error of 0.6\%~\cite{EPJ_Lumi} while the leptonic width 
has been measured by KLOE with a 1.7\% uncertainty \cite{PLB_Gee}. The 
fit has been repeated by changing the value of both quantities of 
$\pm 1\sigma$.

\item {\em Beam energy scale}\\
The beam energy scale also affects the fit due to the explicit \roots\ 
dependence of the theoretical function. Bhabha scattering events allow 
a relative calibration of the energy scale with a precision of 30 keV 
each 100 nb$^{-1}$ of integrated luminosity. 
As mentioned in the introduction, an absolute calibration of 150 keV has 
been applied to match the measured value of $M_{\phi}$. 
The fit to the Dalitz plot has been repeated without applying this 
correction.

\item {\em Photon efficiency}\\
The data-MC correction of the cluster efficiency curve modifies the shape 
of the Dalitz plot.
A different pa\-ra\-me\-tri\-za\-tion of the cluster efficiency curves in 
the Monte Carlo \cite{KN_Ks2pi_2006} has been used to evaluate anew the 
smearing matrix, the analysis efficiency and the background contribution. 
The fit has been repeated in these conditions.

\item {\em $\chi^2$ cut}\\
To test the systematic contribution of the chosen $\chi^2$ cut, we have 
repeated the whole analysis hardening the $\chi^2_{\rm Fit2}/{\rm N_{dof}}$ 
cut from 5 to 3. In this way the event counting is improved due to the 
large reduction of background while the analysis efficiency is not as 
flat as before along the Dalitz plot due to the ISR tails. 

\item {\em Smearing matrix}\\
From the measured quantity of wrong photon pairing 
(Sec.~\ref{Sec:FitMethod}), 
the fraction of off-diagonal events in the smearing matrix has been 
increased by 8\%.

\item {\em Interference}\\
In the standard fit function, the radiative corrections, the analysis 
efficiency and the smearing matrix used in the interference term are 
obtained from an MC sample of \wpi\ events. The fit is repeated by using 
the corresponding quantities estimated with $S\gamma$ events.

\item {\em Background}\\
The scale factors $\alpha_1$, obtained when fitting the back\-ground-enriched 
distributions, are applied to the residual background contamination.

\end{enumerate}

As shown in Tabs.~\ref{Tab:SystKL}, \ref{Tab:SystNS}, the KL fit 
provides stable results when the systematic changes are applied. 
On the other hand, the NS fit shows very large parameter variations 
due to the presence of the continuum background term.

%------------------------------------------------------------------------------
\subsection{Fit results}
%------------------------------------------------------------------------------

\subsubsection{Kaon Loop model}

\begin{figure*}[!t]
  \begin{center}
  \begin{tabular}{cc}
    \hspace{-1cm}
    \resizebox{0.5\textwidth}{!}{\includegraphics{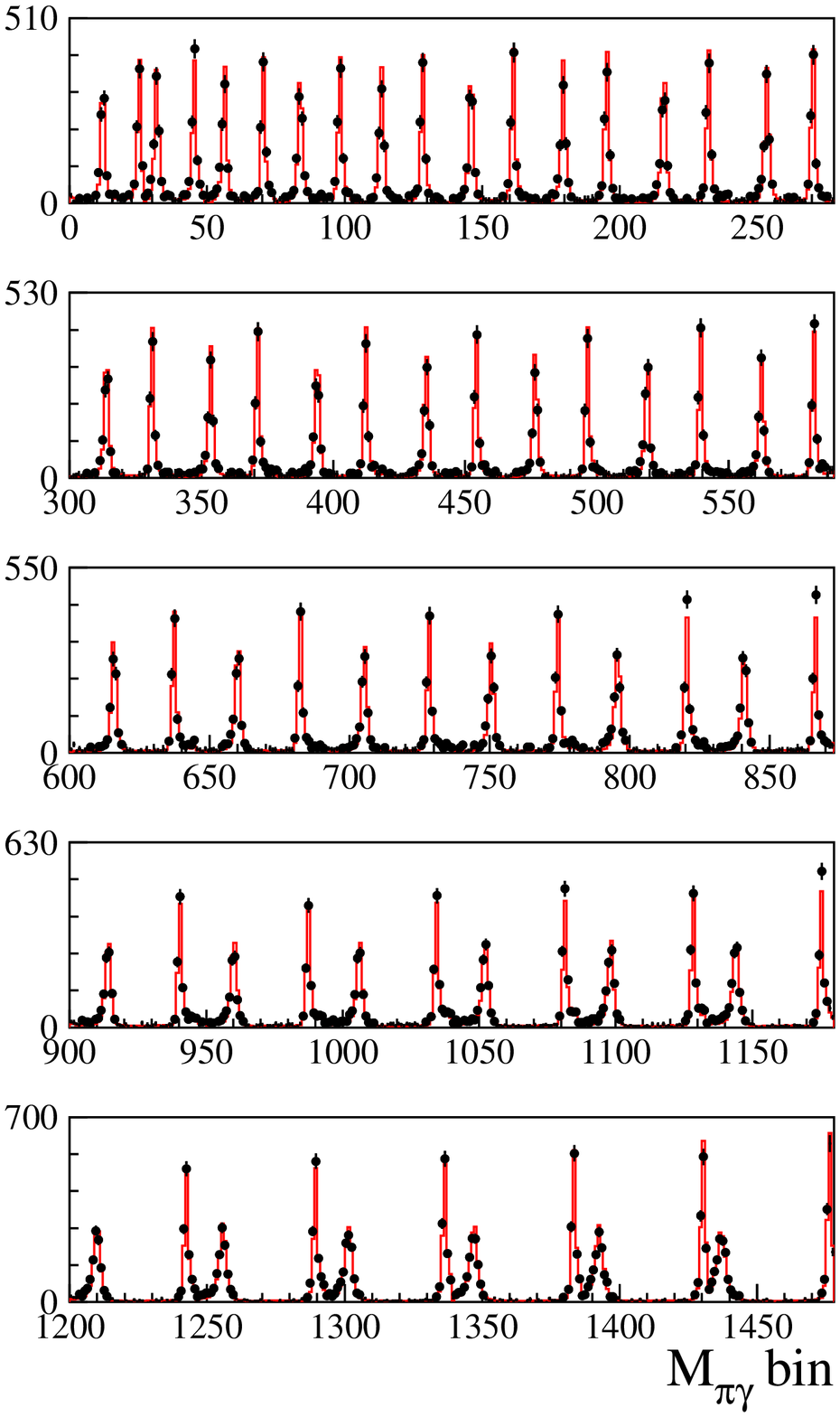}} &
    \hspace{-1cm}
    \resizebox{0.5\textwidth}{!}{\includegraphics{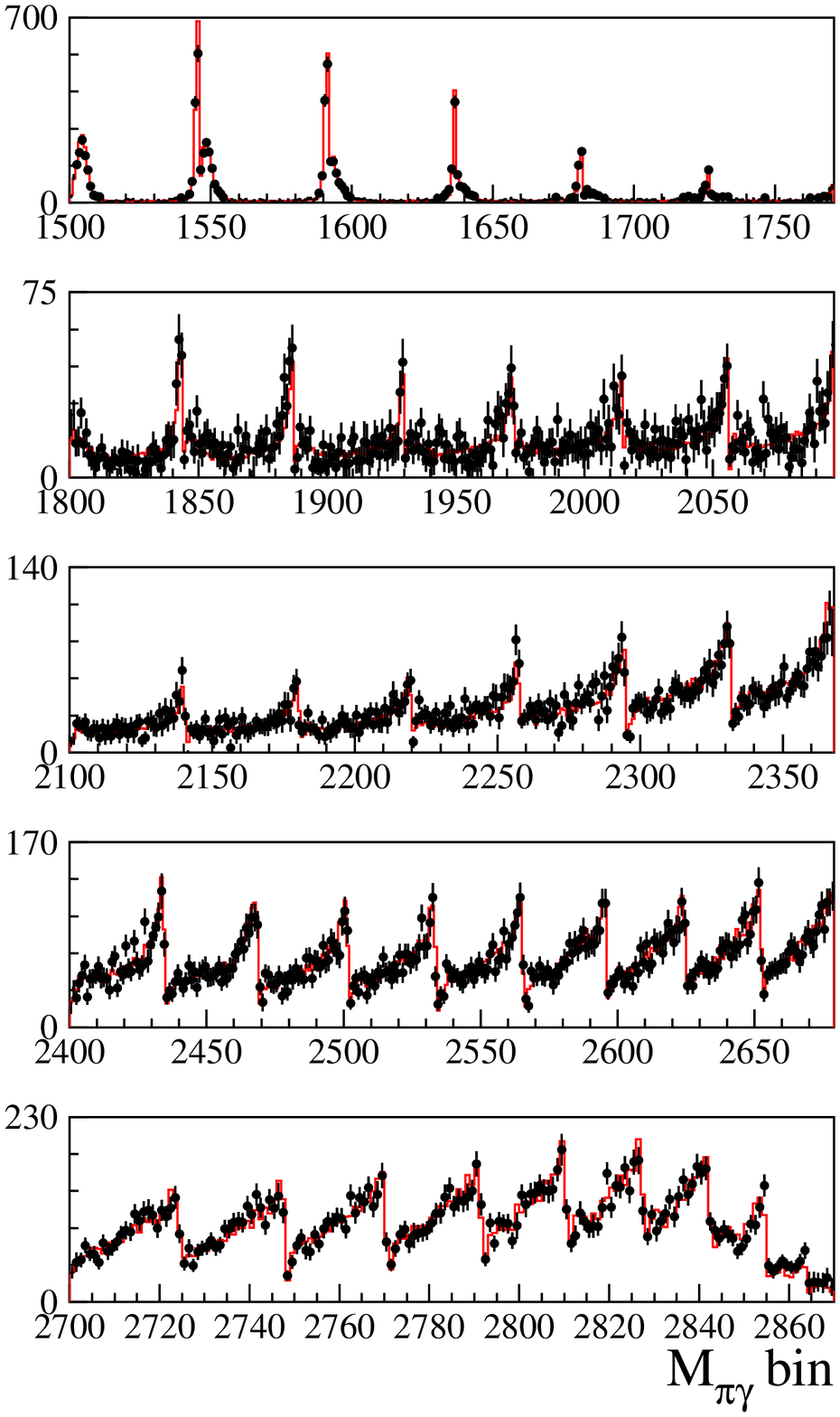}} \\
  \end{tabular}
  \caption{Fit result for Kaon Loop model. Black dots are data while the 
    solid line represents the KL resulting shape.}
  \label{Fig:FitK1}
  \end{center}
\end{figure*}

The Kaon Loop fit results are listed in Tabs.~\ref{Tab:KloopScalars},
\ref{Tab:FitVMD} for the six accepted variants.
In Fig.~\ref{Fig:FitK1} the distributions of the data points for all 
slices of the Dalitz plot with the superimposed fit function of the
variant with the best $P(\chi^2)$ is shown.
Here the \mpg\ {\it vs} \mpp\ histogram is sliced in \mpp\ and for each 
slice the \mpg\ projection of the allowed phase-space region is plotted, 
one after the other. 
To understand the relative importance of the different fitting terms, 
their contributions are shown in Fig.~\ref{Fig:AllShapesFit}.left.
As expected, the VMD ($S\gamma$) term is dominating in the region below 
(above) 700 MeV. The interference term is concentrated around 600 MeV.
As a fit result, we use the central value and the errors from the best 
fit, adding the systematic error discussed in the previous section and
an extra error associated with the theoretical model. This last error 
is evaluated as the maximum variation between the central value obtained 
by the best fit ($P(\chi^2)=0.145$) and the other five accepted fits.
\begin{table*}[!t]
  \caption{Fit results for the scalar term in the Kaon Loop model.
    Different rows are obtained for the accepted sets of parameters 
    from Ref.~\cite{KaonLoopNew}.}
  \label{Tab:KloopScalars}
  \begin{center}
    \renewcommand{\tabcolsep}{5mm}
    \renewcommand{\arraystretch}{0.9}        
    \begin{tabular}{|c|ccc|c|cc|}\hline
      Fit & $M\!\subfo$\,(MeV) & $g_{\!f_{0}K^{+}K^{-}}$\,(GeV) & 
      $g_{\!f_{0}\pi^{+}\pi^{-}}$\,(GeV) & 
      $M_\sigma ,\Gamma\!_\sigma$\,(MeV) &
      $\chi^2$/N$_{\rm dof}$ & $P(\chi^2)$ \\ \hline
      K1 & $976.8\pm 0.3$ & $3.76\pm 0.04$ & $-1.43\pm 0.01$ & 
      $462\,,286$ & 2754/2676 & $0.145$ \\
      K2 & $986.2\pm 0.3$ & $3.87\pm 0.08$ & $-2.03\pm 0.02$ & 
      $485\,,240$ & 2792/2676 & $0.058$ \\
      K3 & $985.2\pm 0.2$ & $4.92\pm 0.06$ & $-1.92\pm 0.01$ & 
      $472\,,320$ & 2809/2676 & $0.036$ \\
      K4 & $982.3\pm 0.4$ & $4.02\pm 0.06$ & $-1.76\pm 0.02$ &
      $415\,,260$ & 2787/2676 & $0.066$ \\
      K5 & $983.3\pm 0.2$ & $3.75\pm 0.02$ & $-1.40\pm 0.01$ & 
      $529\,,366$ & 2823/2676 & $0.024$ \\
      K6 & $986.9\pm 0.1$ & $3.28\pm 0.05$ & $-1.90\pm 0.01$ &
      $566\,,264$ & 2799/2676 & $0.048$ \\ \hline
    \end{tabular}
  \end{center}
\end{table*}
\begin{table*}[!t]
  \caption{Fit results for the VMD parametrization.
    The first six rows are obtained in the KL approach for the accepted 
    sets of parameters from Ref.~\cite{KaonLoopNew} while the last line
    is the result of the NS fit.}
  \label{Tab:FitVMD}
  \begin{center}
    \renewcommand{\tabcolsep}{3mm}
    \renewcommand{\arraystretch}{0.9}    
    \begin{tabular}{|c|ccccccc|}\hline
      Fit & $\alpha_{\rho\pi}$ & $C_{\omega\pi}$ (GeV$^{-2}$) & 
      $\phi_{\omega\pi}$ & $C_{\rho\pi}$ (GeV$^{-2}$) & $\phi_{\rho\pi}$ &
      $\delta_{b_\rho}$ ($^{\circ}$) & $M_{\omega}$ (MeV) \\ \hline
      K1 & $0.58\pm 0.11$ & $0.850\pm 0.010$ & $0.46\pm 0.13$ & 
      $0.260\pm 0.185$ & $3.11\pm 3.12$ & $33.0\pm 9.7$ & $782.52\pm 0.29$ \\
      K2 & $0.68\pm 0.03$ & $0.832\pm 0.003$ & $0.30\pm 0.05$ &
      $0.061\pm 0.211$ & $3.14\pm 3.08$ & $23.6\pm 4.1$ & $782.20\pm 0.11$ \\
      K3 & $0.66\pm 0.17$ & $0.836\pm 0.004$ & $0.33\pm 0.08$ &
      $0.084\pm 0.056$ & $3.14\pm 3.14$ & $25.2\pm 6.2$ & $782.26\pm 0.28$ \\
      K4 & $0.64\pm 0.05$ & $0.836\pm 0.002$ & $0.38\pm 0.06$ &
      $0.061\pm 0.005$ & $3.14\pm 0.04$ & $31.3\pm 2.4$ & $782.28\pm 0.14$ \\
      K5 & $0.62\pm 0.01$ & $0.838\pm 0.006$ & $0.27\pm 0.04$ &
      $0.298\pm 0.126$ & $3.13\pm 0.07$ & $10.4\pm 6.5$ & $782.41\pm 0.07$ \\
      K6 & $0.58\pm 0.04$ & $0.843\pm 0.004$ & $0.30\pm 0.06$ &
      $0.061\pm 0.003$ & $3.14\pm 0.01$ & $33.7\pm 4.8$ & $782.48\pm 0.13$ \\ \hline
      NS & $1.43\pm 0.04$ & $0.953\pm 0.003$ & $0.00\pm 0.01$ & 
      $0.270\pm 0.039$ & $3.11\pm 0.14$ & $73.1\pm 1.6$ & $781.80\pm 0.11$ \\ \hline
    \end{tabular}
  \end{center}
\end{table*}
The extracted parameters are:
\begin{eqnarray}
  && M\!\subfo = ( 976.8 \pm 0.3_{\rm\,fit}\,^{+0.9}_{-0.6}\,_{\rm syst} + 
  10.1_{\rm\,mod} )\ {\rm MeV} \nonumber \\
  && g_{\!f_{0}K^{+}K^{-}} =  
  ( 3.76 \pm 0.04_{\rm\,fit}\,^{+0.15}_{-0.08}\,_{\rm syst}\,
  ^{+1.16}_{-0.48}\,_{\rm mod} )\ {\rm GeV} \nonumber \\
  && g_{\!f_{0}\pi^{+}\pi^{-}} = 
  ( -1.43 \pm 0.01_{\rm\,fit}\,^{+0.01}_{-0.06}\,_{\rm\,syst}\,
  ^{+0.03}_{-0.60}\,_{\rm mod} )\ {\rm GeV} \nonumber \\
  && R\subfo = \frac{g^2_{\!f_{0}K^{+}K^{-}}}{g^2_{\!f_{0}\pi^{+}\pi^{-}}} = 
  6.9 \pm 0.1_{\rm\,fit}\,^{+0.2}_{-0.1}\,_{\rm syst}\,
  ^{+0.3}_{-3.9}\,_{\rm mod} \nonumber \\
  && g_{\phi f_{0}\gamma} = ( 2.78\, ^{+0.02}_{-0.05}\,_{\rm fit} \,
  ^{+0.13}_{-0.05}\,_{\rm syst} + 1.31_{\rm\,mod})\ {\rm GeV}^{-1}\nonumber 
\end{eqnarray}
The first three quantities are the parameters directly extracted from the
fit while the other two are derived. The $g_{\phi f_{0}\gamma}$ coupling
is obtained using the formula

\begin{displaymath}
  g_{\phi f_{0}\gamma} =
  \sqrt{\frac{3}{\alpha}
    \left( \frac{2M_\phi}{M_\phi^2-M_{f_0}^2}\right)^3
    \Gamma_{\phi}\,3\,\br(\phi\to S\gamma\to\pi^0\pi^0\gamma)}.
\end{displaymath}

\subsubsection{No Structure model}

\begin{figure*}[!t]
  \begin{center}
  \begin{tabular}{cc}
    \hspace{-1cm}
    \resizebox{0.5\textwidth}{!}{\includegraphics{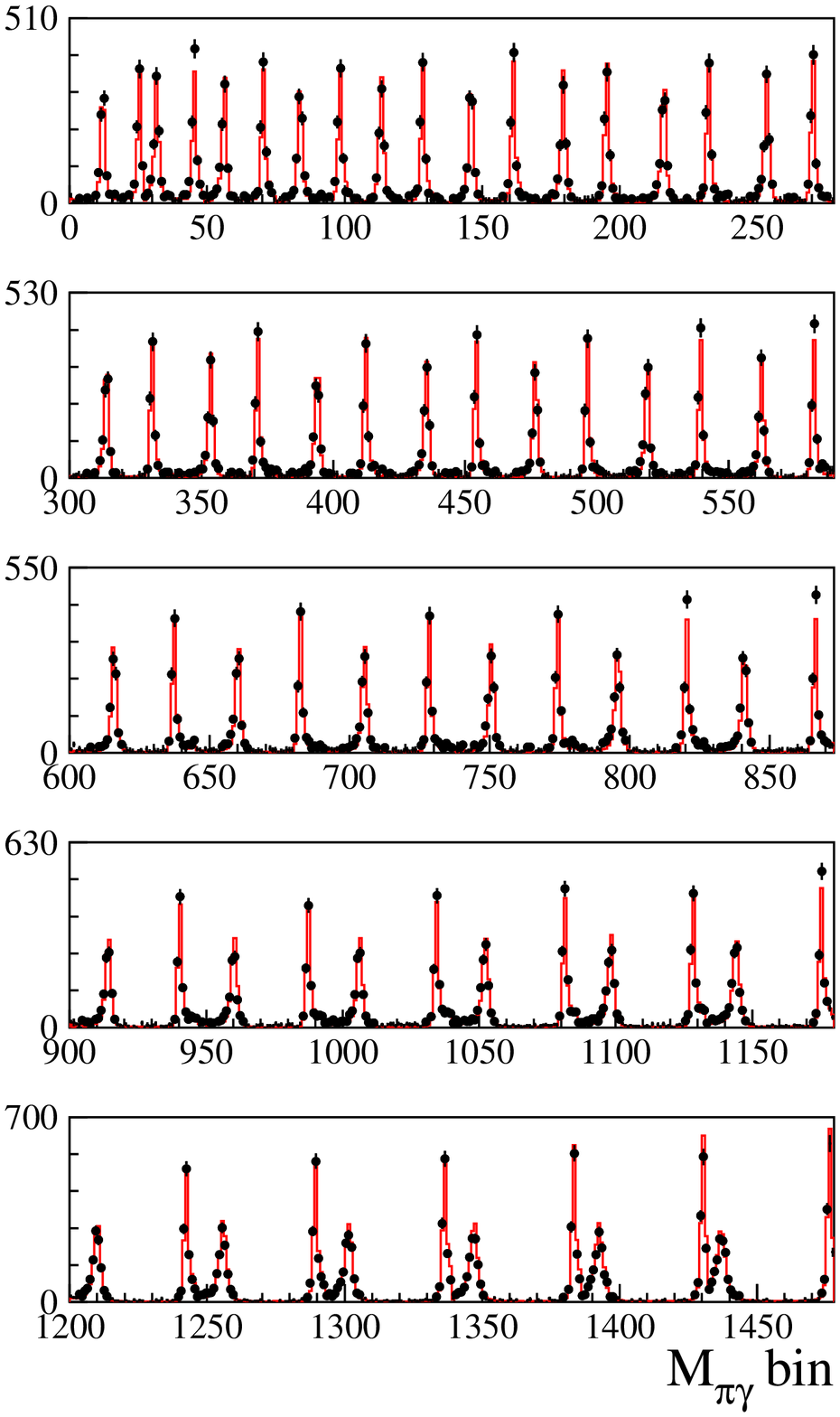}} &
    \hspace{-1cm}
    \resizebox{0.5\textwidth}{!}{\includegraphics{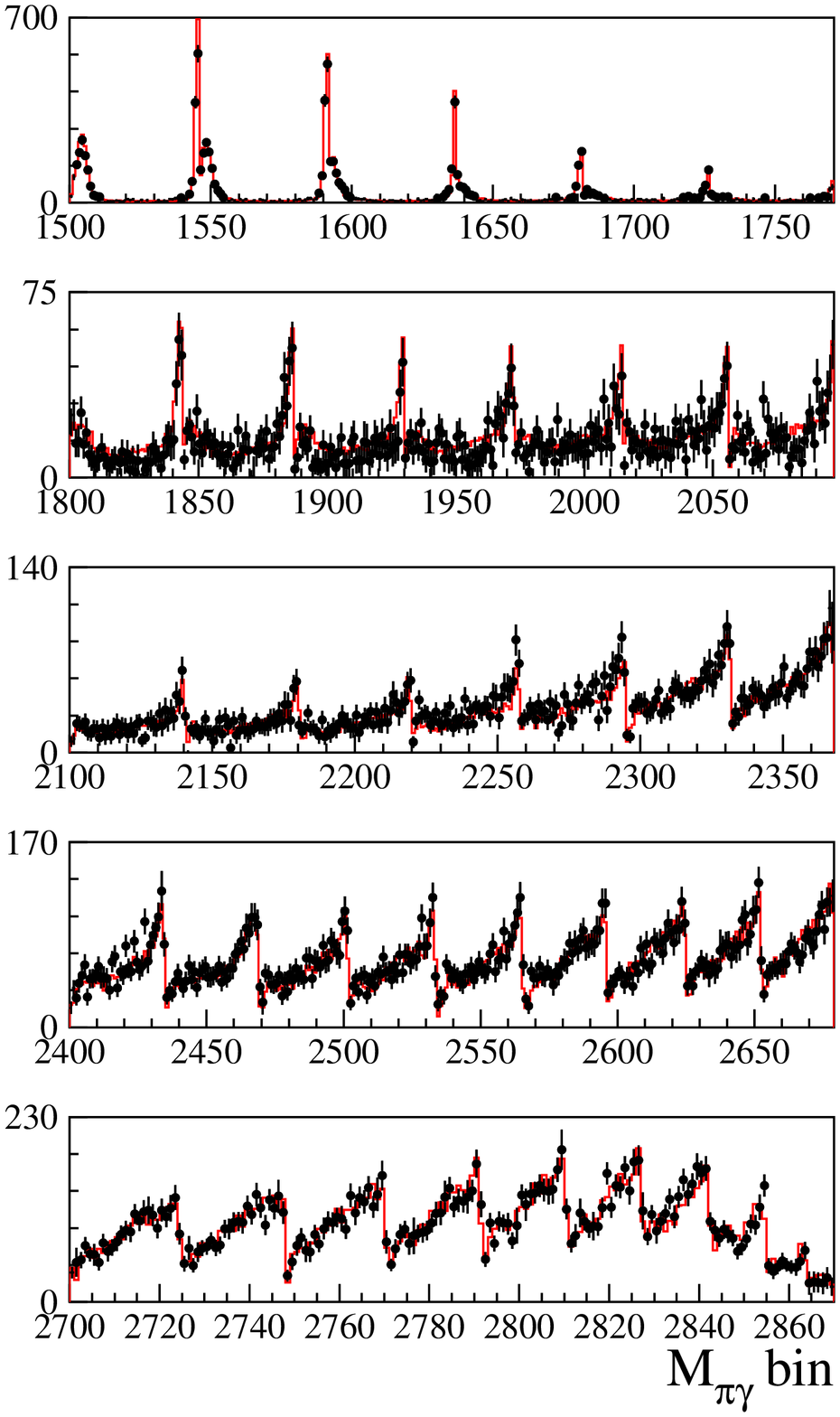}} \\
  \end{tabular}
  \end{center}
  \caption{Fit result for No Structure model. Black dots are data while the 
    solid line represents the NS resulting shape.}
  \label{Fig:FitNS}
\end{figure*}

\begin{figure*}[!t]
  \begin{center}
  \begin{tabular}{cc}
    \resizebox{0.45\textwidth}{!}{\includegraphics{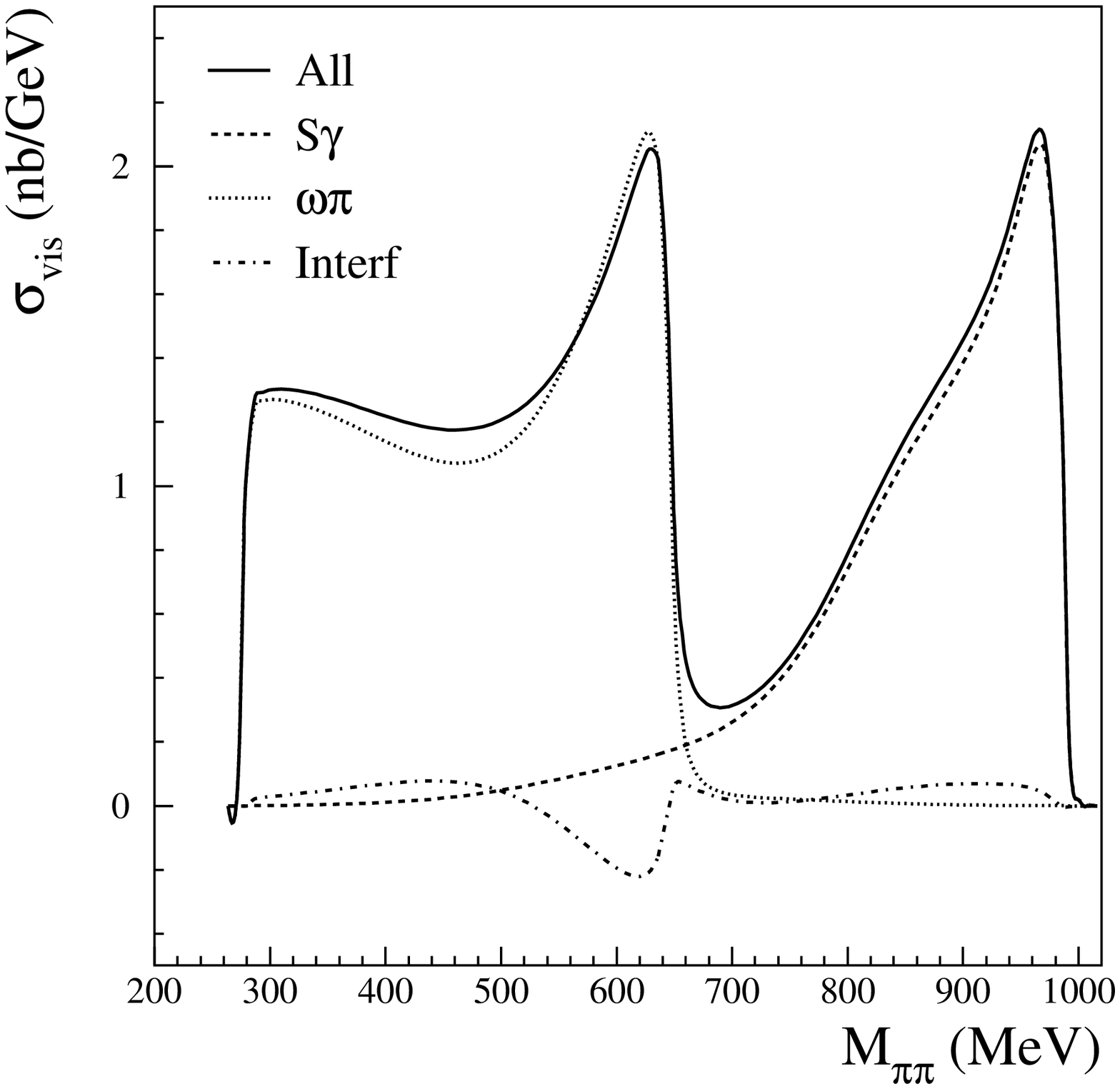}} &
    \resizebox{0.45\textwidth}{!}{\includegraphics{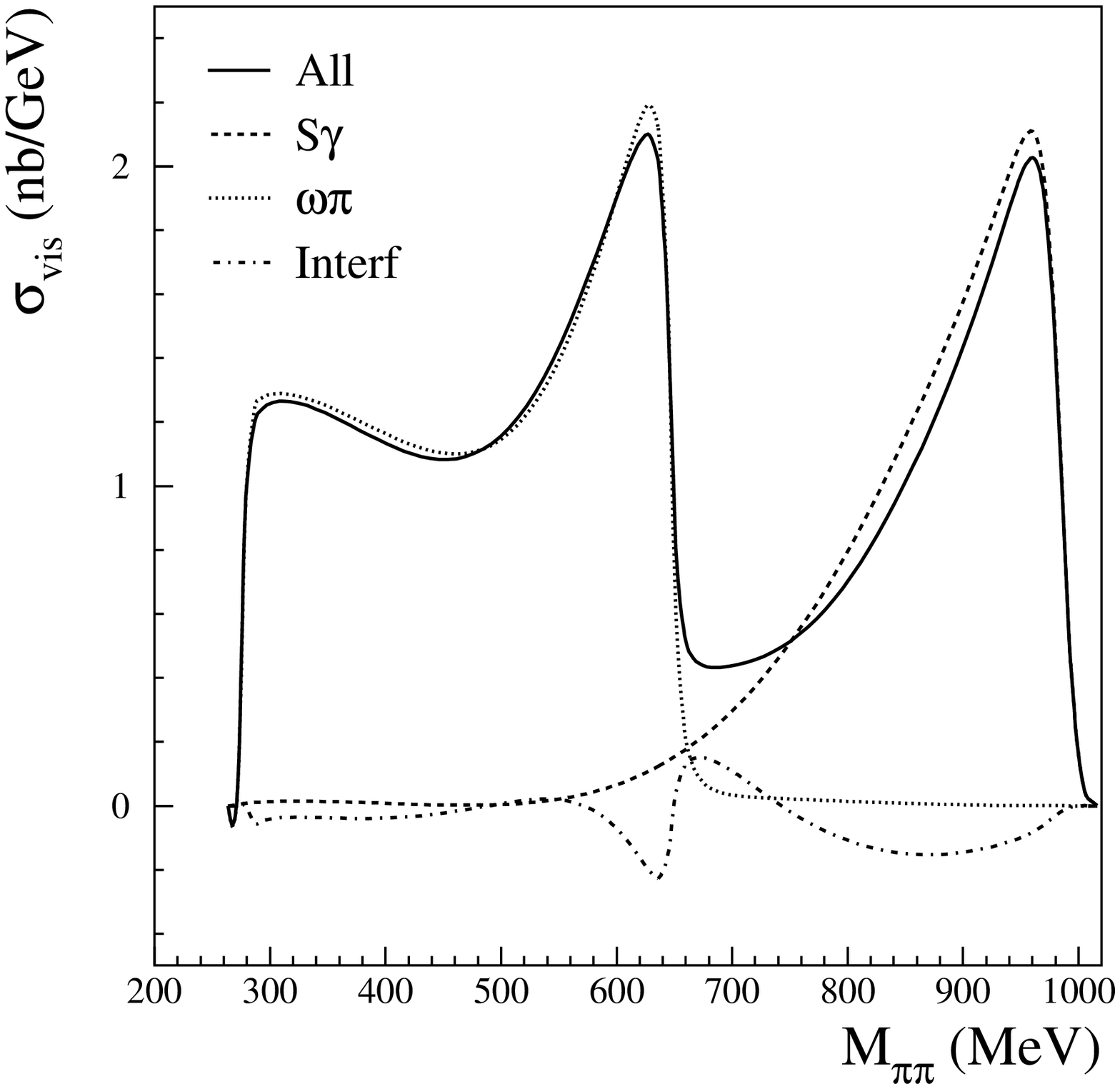}} \\
  \end{tabular}
  \end{center}
  \caption{Resulting $\pio\pio\gamma$ contributions for Kaon Loop (left) 
    and No Structure (right) models.}
  \label{Fig:AllShapesFit}
\end{figure*}

In the No Structure model the resonant term is described with a single 
narrow meson pole, the \fo(980), added to a continuum $\phi\to\pio\pio\gamma$ 
background described by the three free parameters $a_0$, $a_1$ and $b_1$. 
The fit quality, having a $P(\chi^2)=0.042$, is a little worse than 
the best KL result but still acceptable (see Fig.~\ref{Fig:FitNS}). 
The different components of the fit 
are shown in Fig.~\ref{Fig:AllShapesFit}.right. Again, for 
$\mpp > 700$~MeV the scalar contribution is clearly dominant. However, 
contrary to the KL case, the interference term gets negative in this 
region, so that the scalar term is slightly increased. 
The VMD fit parameters are listed in Tab.~\ref{Tab:FitVMD} while for 
the scalar term we have:
\begin{eqnarray}
  && M\!\subfo =  
  ( 984.7 \pm 0.4_{\rm\,fit}\, ^{+2.4}_{-3.7}\,_{\rm\,syst} )\ {\rm MeV} \nonumber \\
  && g_{\!f_{0}K^{+}K^{-}} =  
  ( 0.40 \pm 0.04_{\rm\,fit}\, ^{+0.62}_{-0.29}\,_{\rm\,syst} )\ {\rm GeV} \nonumber \\
  && g_{\!f_{0}\pi^{+}\pi^{-}} = 
  ( 1.31 \pm 0.01_{\rm\,fit}\, ^{+0.09}_{-0.03}\, _{\rm\,syst} )\ {\rm GeV} \nonumber \\
  && R\subfo = \frac{g^2_{\!f_{0}K^{+}K^{-}}}{g^2_{\!f_{0}\pi^{+}\pi^{-}}} = 
  0.09 \pm 0.02_{\rm\,fit}\,^{+0.44}_{-0.08}\,_{\rm\,syst} \nonumber \\
  && g_{\phi f_0\gamma} = 
  ( 2.61 \pm 0.02_{\rm\,fit}\, ^{+0.31}_{-0.08}\,_{\rm\,syst} )\ 
  {\rm GeV^{-1}} \nonumber \\
  && a_0 = \hphantom{-}4.19 \pm 0.01_{\rm\,fit}\,^{+1.44}_{-0.42}\,_{\rm syst} \nonumber \\
  && a_1 = \hphantom{-}1.31 \pm 0.01_{\rm\,fit}\,^{+1.60}_{-0.34}\,_{\rm syst} \nonumber \\
  && b_1 = -1.50 \pm 0.02_{\rm\,fit}\,^{+0.86}_{-0.42}\,_{\rm\,syst}\nonumber 
\end{eqnarray}
Here the only parameter that is not directly extracted from the fit is
R\subfo.


\subsubsection{KL, NS combined results}

\begin{figure}[!t]
    \resizebox{\columnwidth}{!}{\includegraphics{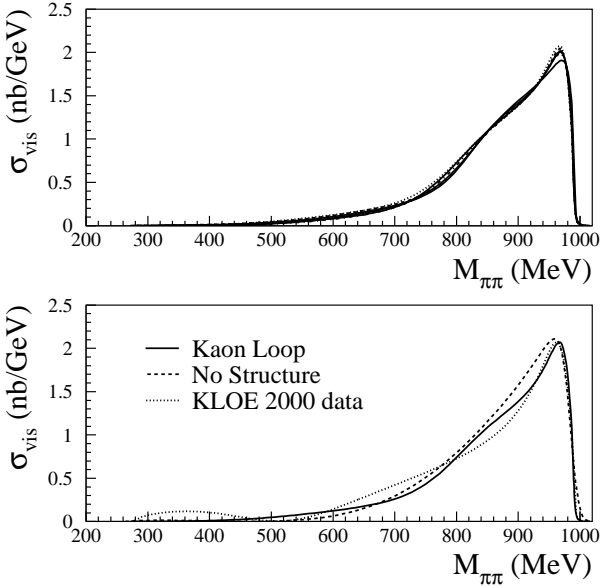}}
  \caption{\mpp\ distribution of the $\phi\to S\gamma\to\pio\pio\gamma$ 
    process obtained from the fit. 
    Top: comparison between the six accepted results of the KL model. 
    Bottom: comparison between the best KL fit, the NS resulting shape
    and the previous KLOE measurement~\cite{PLBf0n2002}.}
  \label{Fig:ScalarShapeFit}
\end{figure}

In Fig.~\ref{Fig:ScalarShapeFit} the comparison of the scalar contributions 
obtained from the six KL fit results shows stable resulting shapes.
On the contrary, comparing the KL and NS curves there are
differences at a level of few \% (Fig.~\ref{Fig:ScalarShapeFit}), 
mainly due to the interference term. In the same figure, the result of the 
fit obtained with the old \fo(980)+$\sigma(600)$ parametrization on 2000 data 
\cite{PLBf0n2002} is shown. The small bump below 500 MeV is now described 
by the $S\gamma$-VMD interference.
By integrating the KL, NS distributions and normalizing to the $\phi$ 
production cross section, an effective \br\ for the 
$\phi\to S\gamma\to\pio\pio\gamma$ process is extracted:
\begin{eqnarray}
  && \br(\phi\to S\gamma\to\pio\pio\gamma) = \nonumber \\ \nonumber
  && \ \ \ \ \ \ \ ( 1.07\,^{+0.01}_{-0.03}\,_{\rm fit}\, 
  ^{+0.04}_{-0.02}\,_{\rm syst}  
  \;^{+0.05}_{-0.06}\;_{\rm mod}) \times 10^{-4}
\end{eqnarray}
The central value is given by the KL model with the best $P(\chi^2)$,
the fit error has been evaluated as the maximum excursion obtained when 
varying the fit errors 
by $\pm 1\sigma$ and the model error corresponds to the maximum variation 
of the central value with respect to the other five accepted fit results 
of the KL model and to the NS description.

\subsubsection{Extrapolation to other \roots\ points}

\begin{figure*}[!t]
  \begin{center}
    \begin{tabular}{cc}
      \resizebox{0.8\columnwidth}{!}{\includegraphics{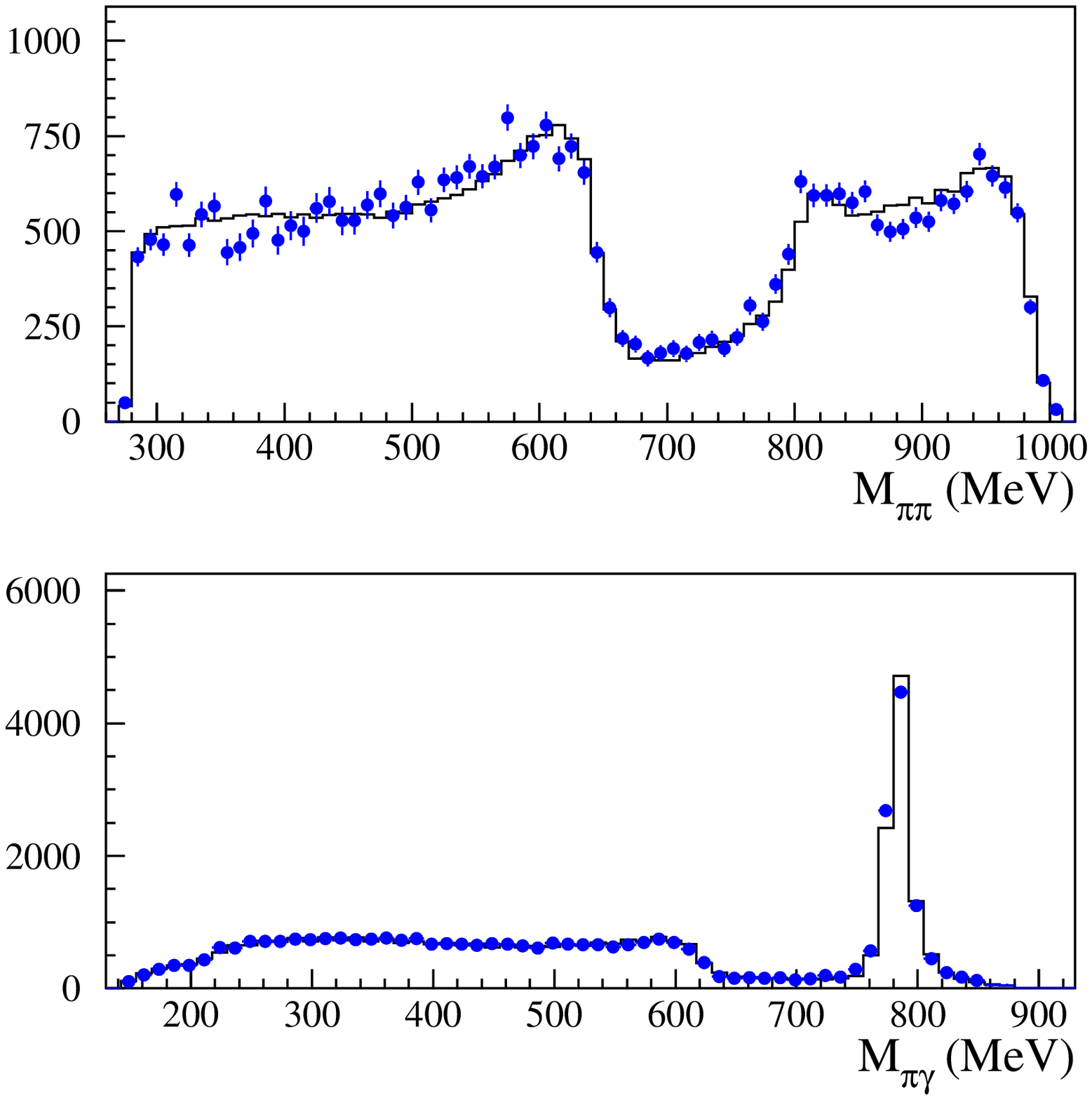}} &
      \resizebox{0.8\columnwidth}{!}{\includegraphics{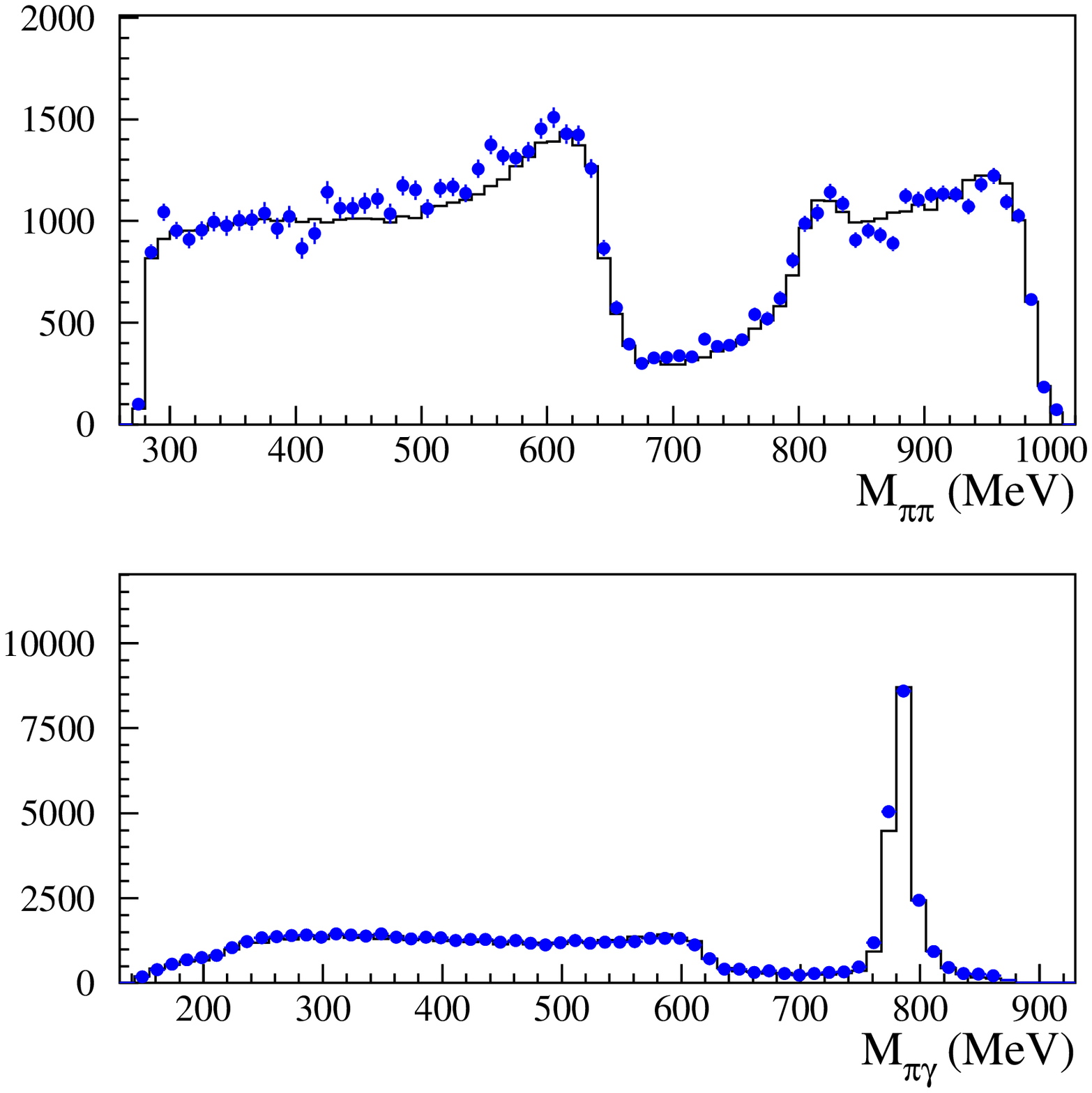}} \\
      \resizebox{0.8\columnwidth}{!}{\includegraphics{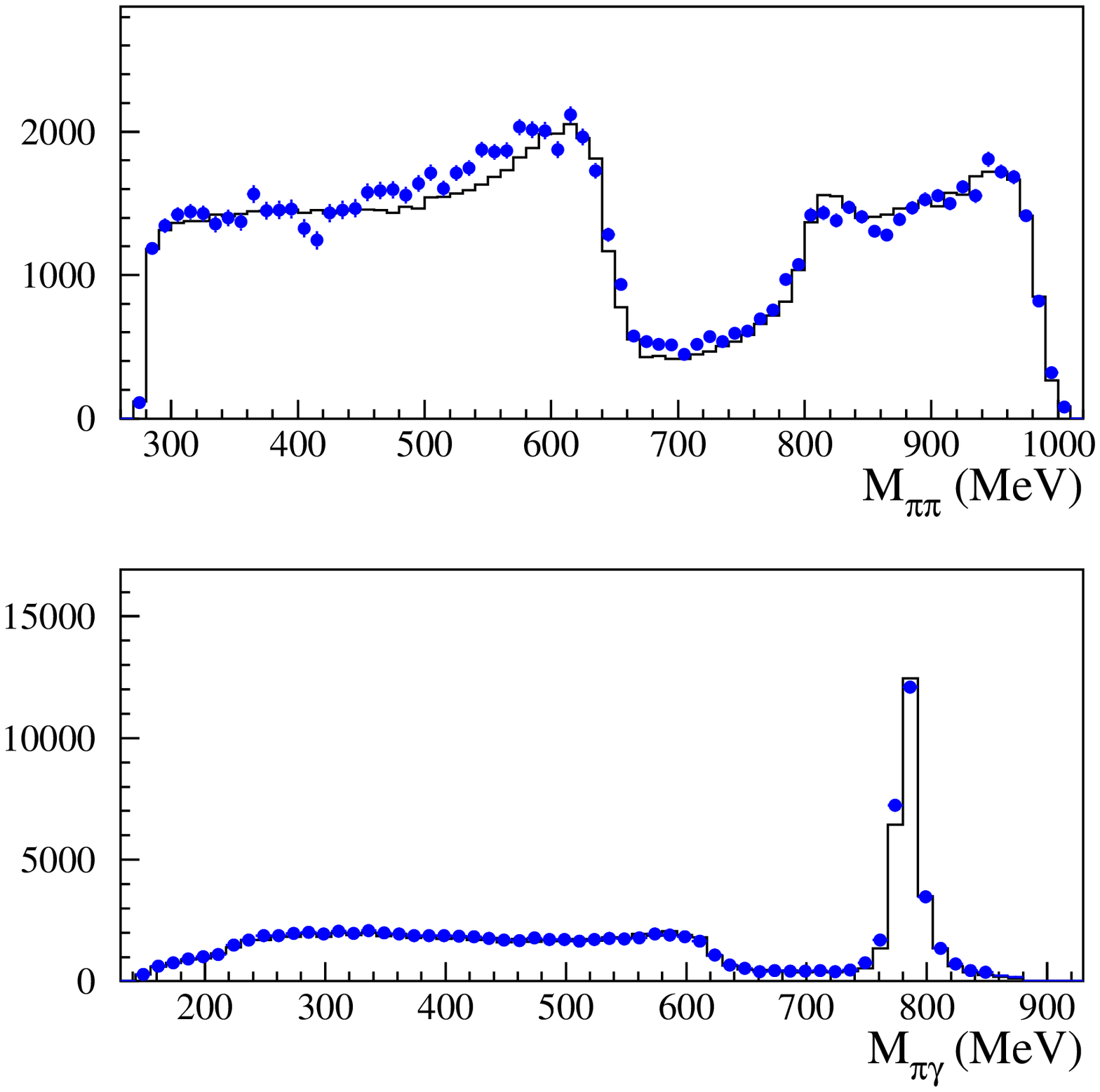}} &
      \resizebox{0.8\columnwidth}{!}{\includegraphics{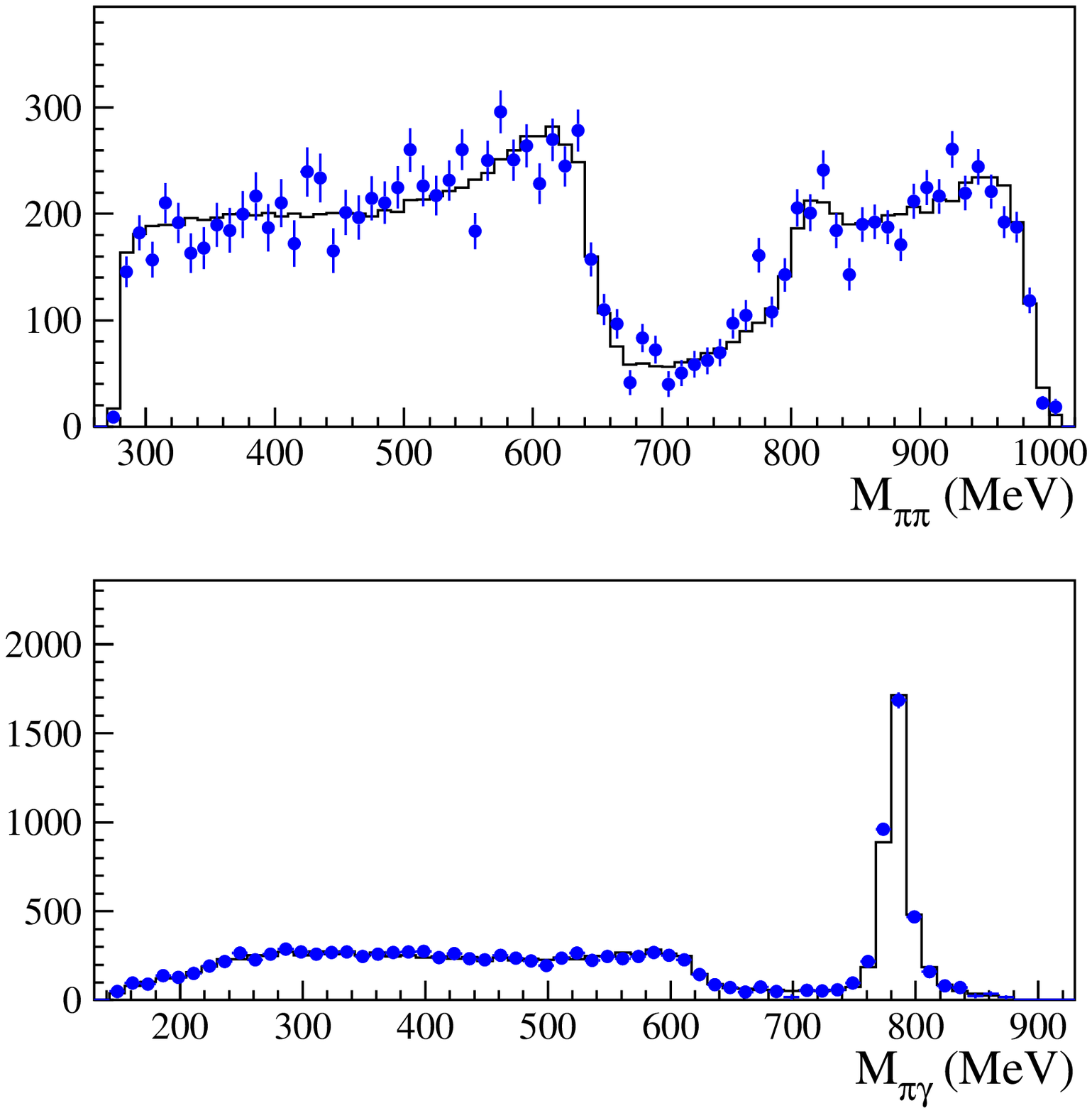}} \\
    \end{tabular}
  \end{center}
  \caption{Dalitz plot projections for runs with different \roots: 
    1019.55 MeV (top-left), 1019.65 MeV (top-right), 1019.85 MeV
    (bottom-left) and 1019.95 MeV (bottom-right). Data are reported 
    in dots while the solid line represents the expected shape 
    extrapolated from fit K1 of the Kaon Loop model.}
  \label{Fig:KLNear}
\end{figure*}

\begin{figure*}
  \begin{center}
    \begin{tabular}{cc}
      \resizebox{0.8\columnwidth}{!}{\includegraphics{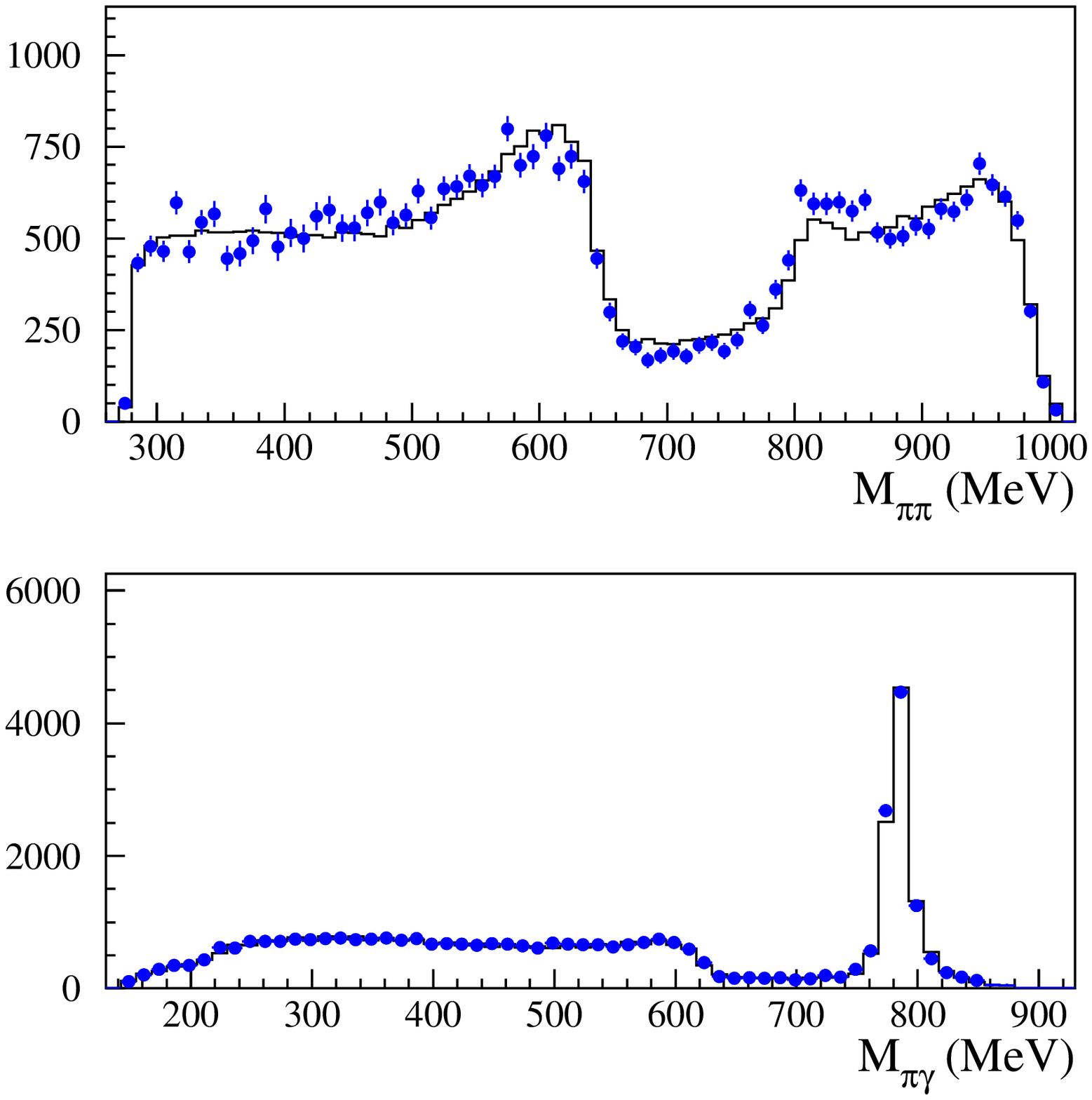}} &
      \resizebox{0.8\columnwidth}{!}{\includegraphics{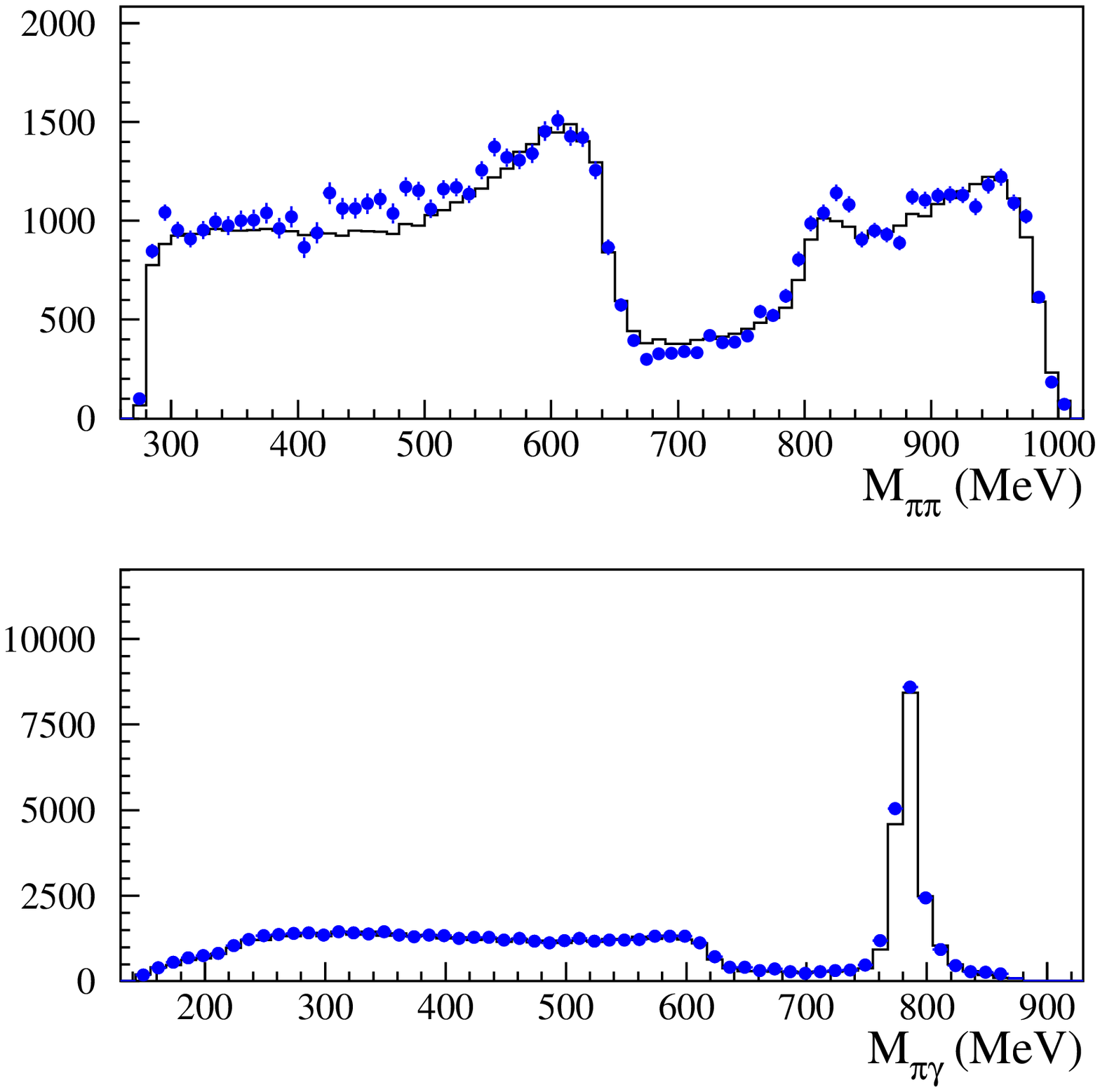}} \\
      \resizebox{0.8\columnwidth}{!}{\includegraphics{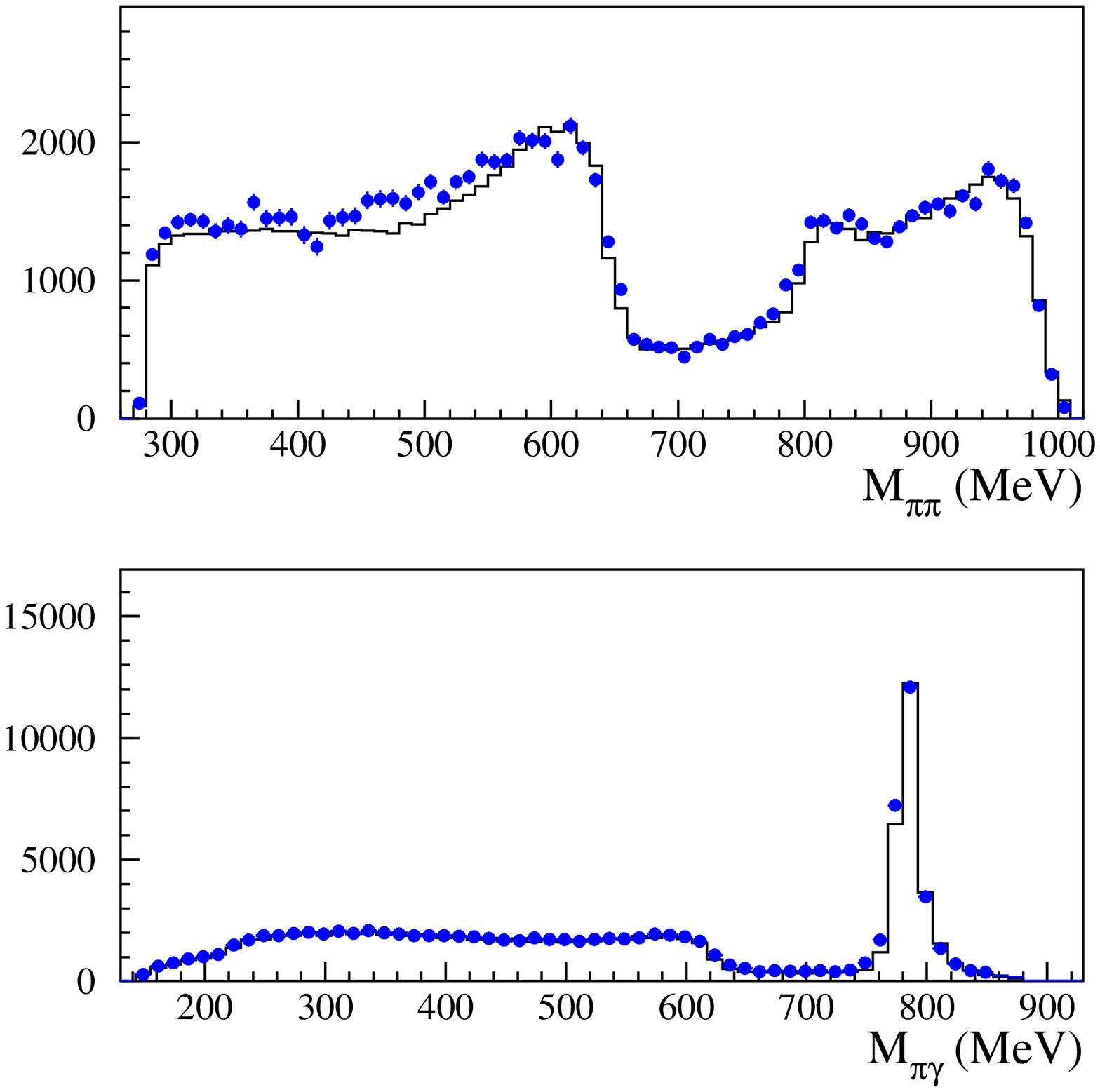}} &
      \resizebox{0.8\columnwidth}{!}{\includegraphics{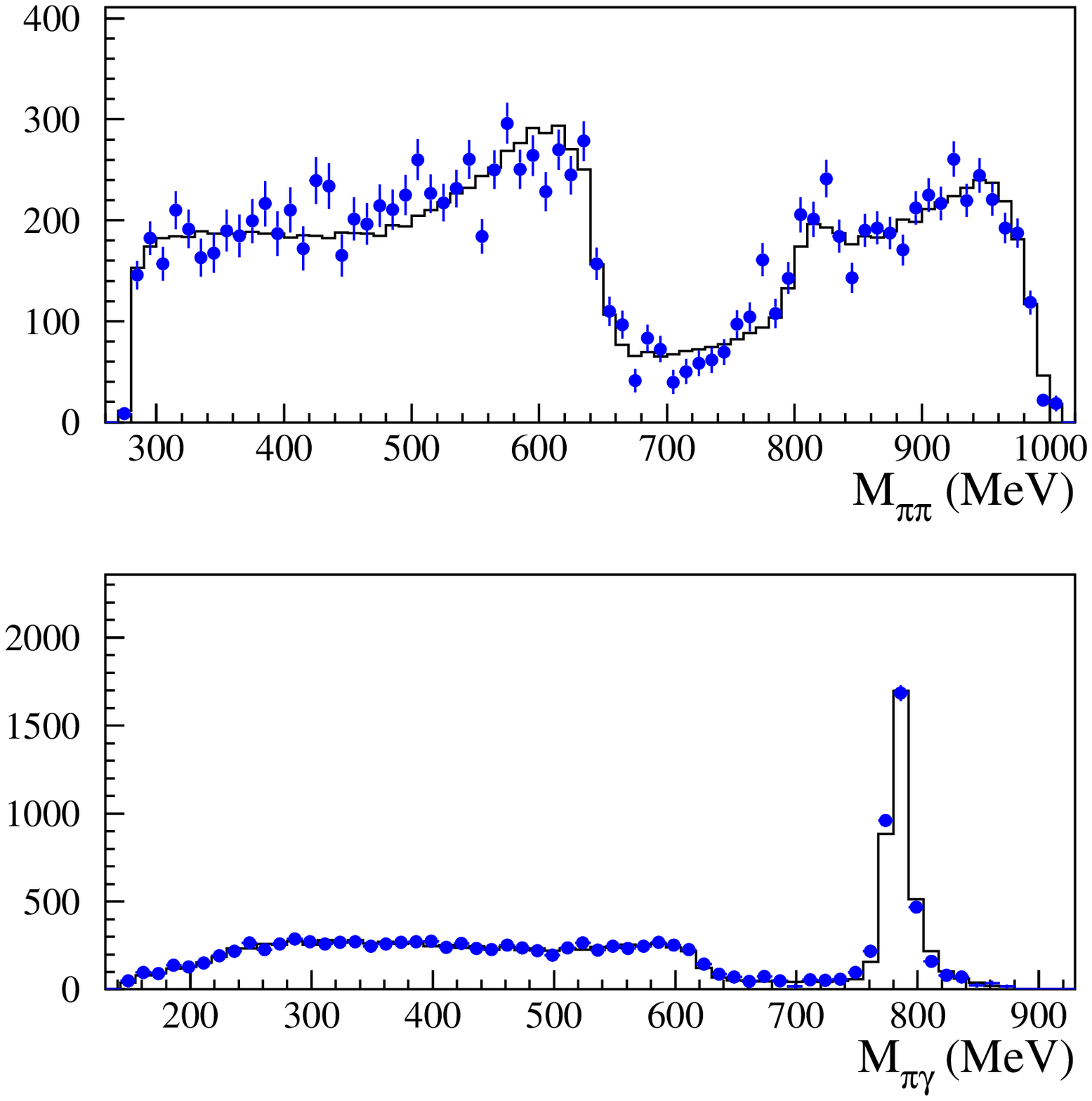}} \\
    \end{tabular}
  \end{center}
  \caption{Dalitz plot projections for runs with different \roots: 
    1019.55 MeV (top-left), 1019.65 MeV (top-right), 1019.85 MeV
    (bottom-left) and 1019.95 MeV (bottom-right). Data are reported 
    in dots while the solid line represents the expected shape 
    extrapolated from NS fit.}
  \label{Fig:NSNear}
\end{figure*}

As a last check, we extrapolate both KL and NS fit results to the 
four closest \roots\ points: 1019.55 MeV (42 pb$^{-1}$), 1019.65 MeV 
(77 pb$^{-1}$), 1019.85 MeV (100 pb$^{-1}$) and 1019.95 MeV (15 pb$^{-1}$), 
scaling for the integrated luminosity. As shown in Figs.~\ref{Fig:KLNear} 
and \ref{Fig:NSNear}, a good agreement is obtained in all the \mpp\ and
\mpg\ distributions.

%------------------------------------------------------------------------------
\subsection{\boldmath Comparison with $\pip\pim\gamma$ final state}
%------------------------------------------------------------------------------

In principle it is possible to compare these results with those obtained
by KLOE on the charged channel $\ep\el\to\pip\pim\gamma$ \cite{PLBf0c2005}.
However in this case the non resonant background (dominated by the $\rho$
radiative tail) is much more important, so that the extraction of the
\fo(980) signal is more difficult and there is a reduced sensitivity to the
presence of the $\sigma(600)$. 

The results of the fit with the KL or NS model yield consistent values 
for the \fo(980) mass and for the branching ratio while large discrepancies 
are observed on the differential cross section, especially for the NS model.

More precisely, by scaling by a factor two the extracted cross-section 
for the $\pi^0\pi^0\gamma$ case, we estimate 
BR($\phi\to S\gamma\to\pip\pim\gamma)=(2.14 \pm 0.14)\times 10^{-4}$,
to be compared to the values from $\pip\pim\gamma$ data:
$(2.1 \pm 0.4)\times 10^{-4}$ (KL model) and $(2.4 \pm 0.5)\times 10^{-4}$ 
(NS model).

For the KL model, a still acceptable agreement on the $f_0(980)$ 
coupling constants is observed, although the $\sigma(600)$ is not 
needed in the $\pip\pim\gamma$ case. This is mainly due to the 
improved parametrization of the KL which had not been used for the 
charged case. For the NS model instead large differences are found
on all couplings.

%==============================================================================
\section{Conclusions}
%==============================================================================

The analysis of the $\pio\pio\gamma$ final state presented here treats 
equally the two main production mechanisms, the VMD and $S\gamma$ processes.
The high  statistics (145 pb$^{-1}$ in a single $\sqrt{s}$ bin)  
allowed us to fit with two different theoretical models 
the Dalitz plot distribution. In the Kaon Loop model, the two low mass 
scalars $f_0(980)$ and $\sigma(600)$ are required to adequately fit the 
data. In the case of the No Structure model, we instead find an acceptable 
fit with the $f_0(980)$ meson alone. In the latter formulation 
the low mass \mpp\ behaviour is described by the three free parameters $a_0$, 
$a_1$ and $b_1$ representing the continuum background.
For both models, the resulting fit curve reproduces also the mass
spectrum of all other \roots\ bins around $M_\phi$.

A stable branching ratio of the $\phi\to\pio\pio\gamma$ process is obtained:
\begin{eqnarray} 
  && \br(\phi\to S\gamma\to\pio\pio\gamma) = \nonumber \\ \nonumber
  && \ \ \ \ \ \ \ ( 1.07\,^{+0.01}_{-0.03}\,_{\rm fit}\, 
  ^{+0.04}_{-0.02}\,_{\rm syst}  
  \;^{+0.05}_{-0.06}\;_{\rm mod}) \times 10^{-4} 
\end{eqnarray}
The last error reflects the maximum variation observed when changing 
the fit model. This result is consistent with our previous published 
measurement.

The extracted couplings show that the Kaon Loop model provides a stable 
description of the data with large coupling of \fo(980) to kaons, as also 
indicated by the study of the
$\pip\pim\gamma$ final state. Therefore, these results add evidences 
to a 4-quark structure of the $f_0(980)$ meson. On the other hand, in the
fit with the No Structure model, the $f_0(980)$ coupling to kaons get
substantially reduced with respect to what found with the $\pip\pim\gamma$
channel. However, the 
physical interpretation is more difficult due to the presence of the 
continuum background which differs substantially in the \pio\pio\ 
and \pip\pim\ cases.

%==============================================================================
\section*{Acknowledgements}
%==============================================================================

We have to thank many people who helped us in this study.
We are in debt with N.~N.~Achasov and A.~V.~Kiselev for 
many clarifications on the Kaon Loop model and for the VMD description.
We also acknowledge warmly G.~Isidori, L.~Maiani and S.~Pacetti for 
many fruitful discussions.

We thank the DAFNE team for their efforts in maintaining low background 
running conditions and their collaboration during all data-taking. 
We want to thank our technical staff: 
G.F.Fortugno for his dedicated work to ensure an efficient operation of 
the KLOE Computing Center; 
M.Anelli for his continuous support to the gas system and the safety of
the detector; 
A.Balla, M.Gatta, G.Corradi and G.Papalino for the maintenance of the
electronics;
M.Santoni, G.Paoluzzi and R.Rosellini for the general support to the
detector; 
C.Piscitelli for his help during major maintenance periods.
This work was supported in part by DOE grant DE-FG-02-97ER41027; 
by EURODAPHNE, contract FMRX-CT98-0169; 
by the German Federal Ministry of Education and Research (BMBF) contract 
06-KA-957; 
by Graduiertenkolleg `H.E. Phys. and Part. Astrophys.' of Deutsche 
Forschungsgemeinschaft, Contract No. GK 742; 
by INTAS, contracts 96-624, 99-37
and by the EU Integrated Infrastructure
Initiative HadronPhysics Project under contract number
RII3-CT-2004-506078.

\appendix
%==============================================================================
\section{\boldmath
  Differential cross section at $\roots\simeq M_\phi$}
%==============================================================================
\label{App:Xsec}

The double differential $\pio\pio\gamma$ cross section can be written as 
the sum of three terms: the scalar contribution (proportional to the 
amplitude $|M_{S\gamma}|^2$), the VMD term and their relative interference 
\cite{AchasovPrivateVMD}:
\begin{eqnarray}
  && \frac{d\sigma}{dM_{\pi\pi}dM_{\pi\gamma}} =  
  \frac{\alpha\, M_{\pi\gamma} M_{\pi\pi}} {3(4\pi)^2 s^3} \, 
  \left\{ \frac{g_{\phi\gamma}}{|D_{\phi}(s)|^2}|M_{S\gamma}|^2 +  \right.  
                                                             \\ 
  && \ \ \ \frac{1}{16} F_1(M_{\pi\pi}^2,M_{\pi\gamma}^2) 
  |G_{\rho,\omega}(s,M_{\pi\gamma}^2)|^2 +                 \nonumber \\
  && \ \ \ \frac{1}{16} F_1(M_{\pi\pi}^2,\tilde{M}_{\pi\gamma}^2) 
  |G_{\rho,\omega}(s,\tilde{M}_{\pi\gamma}^2)|^2 +           \nonumber \\
  && \ \ \ \frac{1}{8}  F_2(M_{\pi\pi}^2,M_{\pi\gamma}^2) 
  \,\Re e \,[ G_{\rho,\omega}(s,M_{\pi\gamma}^2) 
  G^*_{\rho,\omega}(s,\tilde{M}_{\pi\gamma}^2)] \pm          \nonumber \\
  && \ \ \ \frac{1}{\sqrt{2}}\, \Re e \,\left[ \frac{g_{\phi\gamma}}{D_{\phi}(s)} 
    M_{S\gamma} 
    [ F_3(M_{\pi\pi}^2,M_{\pi\gamma}^2) G^*_{\rho,\omega}(s,M_{\pi\gamma}^2) + 
    \nonumber \right. \\
  && \ \ \   \left. F_3(M_{\pi\pi}^2,\tilde{M}_{\pi\gamma}^2) 
    G^*_{\rho,\omega}(s,\tilde{M}_{\pi\gamma}^2)] \right] \nonumber
\label{Eq:XsecAll}
\end{eqnarray}
where 
$D_{\phi}(s)$ is the $\phi$ inverse propagator and $g_{\phi\gamma}$ is the 
coupling of the $\phi$ to \ep\el. The general expression for a vector meson
$V$ is $g_{V\gamma} = \sqrt{3 M_V^3 \Gamma_V B(V\to \ep\el)/\alpha}$.
The VMD parametrization contains two terms due to the exchange of identical 
pions ($M_{\pi\gamma}$ {\it vs} $\tilde{M}_{\pi\gamma}$) and their 
interference term. The full expression of the three coefficients 
$F_i(m,m_{\pi\gamma}^2)$ is the following:
\begin{eqnarray}
  && F_1(m^2,m_{\pi\gamma}^2) =
                  m_{\pi_0}^8
        + 2\, m^2 m_{\pi_0}^4 m_{\pi\gamma}^2
        - 4\,     m_{\pi_0}^6 m_{\pi\gamma}^2               \\ & & 
  \ \ \ + 2\, m^4             m_{\pi\gamma}^4
        - 4\, m^2 m_{\pi_0}^2 m_{\pi\gamma}^4
        + 6\,     m_{\pi_0}^4 m_{\pi\gamma}^4    \nonumber  \\ & & 
  \ \ \ + 2\, m^2             m_{\pi\gamma}^6
        - 4\,     m_{\pi_0}^2 m_{\pi\gamma}^6 
        +                     m_{\pi\gamma}^8 
        - 2\,     m_{\pi_0}^6                 s   \nonumber \\ & &
  \ \ \ - 2\, m^2 m_{\pi_0}^2 m_{\pi\gamma}^2 s
        + 2\,     m_{\pi_0}^4 m_{\pi\gamma}^2 s
        - 2\, m^2             m_{\pi\gamma}^4 s   \nonumber \\ & &
  \ \ \ + 2\,     m_{\pi_0}^2 m_{\pi\gamma}^4 s
        - 2\,                 m_{\pi\gamma}^6 s
        +         m_{\pi_0}^4                 s^2 
        +                     m_{\pi\gamma}^4 s^2 \nonumber
\end{eqnarray}
\begin{eqnarray}
  && F_2(m^2,m_{\pi\gamma}^2) =
                  m_{\pi_0}^8
        -     m^6             m_{\pi\gamma}^2
        + 2\, m^4 m_{\pi_0}^2 m_{\pi\gamma}^2               \\ & & 
  \ \ \ + 2\, m^2 m_{\pi_0}^4 m_{\pi\gamma}^2
        - 4\,     m_{\pi_0}^6 m_{\pi\gamma}^2
        - 4\, m^2 m_{\pi_0}^2 m_{\pi\gamma}^4     \nonumber \\ & &
  \ \ \ + 6\,     m_{\pi_0}^4 m_{\pi\gamma}^4 
        + 2\, m^2             m_{\pi\gamma}^6 
        - 4\,     m_{\pi_0}^2 m_{\pi\gamma}^6 
        +                     m_{\pi\gamma}^8     \nonumber \\ & &
  \ \ \ +     m^2 m_{\pi_0}^4                 s
        - 2\,     m_{\pi_0}^6                 s
        + 2\, m^4             m_{\pi\gamma}^2 s 
        - 4\, m^2 m_{\pi_0}^2 m_{\pi\gamma}^2 s   \nonumber \\ & &
  \ \ \ + 2\,     m_{\pi_0}^4 m_{\pi\gamma}^2 s
        -     m^2             m_{\pi\gamma}^4 s 
        + 2\,     m_{\pi_0}^2 m_{\pi\gamma}^4 s
        - 2\,                 m_{\pi\gamma}^6 s   \nonumber \\ & &
  \ \ \ -         m_{\pi_0}^4                 s^2
        -     m^2             m_{\pi\gamma}^2 s^2
        + 2\,     m_{\pi_0}^2 m_{\pi\gamma}^2 s^2
        +                     m_{\pi\gamma}^4 s^2 \nonumber
\end{eqnarray}
\begin{equation}
  F_3(m^2,m_{\pi\gamma}^2) = 
  \frac{(m_{\pi\gamma}^2-m_{\pi_0}^2)^2 s - (s-m^2)^2 m_{\pi\gamma}^2}{s-m^2}
\end{equation}

The quantity $G_{\rho,\omega}(s,M_{\pi\gamma}^2)$ is given by:
\begin{eqnarray}
 &&  G_{\rho,\omega}(s,M_{\pi\gamma}^2) = 
 \frac{ C_{\omega\pi}}{D_{\omega}(M_{\pi\gamma}^2)} + \\
 && \ \ \ \ \ \left(\frac{e^{i\phi_{\omega\phi}(M_{\phi}^2)}\,g_{\phi\gamma}\,
        g_{\phi\rho\pi}\,g_{\rho\pi\gamma}}{D_{\phi}(s)}+C_{\rho\pi} \right) 
  \frac{e^{i\delta_{b_\rho}}}{D_{\rho}(M_{\pi\gamma}^2)} \nonumber
\end{eqnarray}
The first term in the parenthesis is the only resonant component and
includes the $\phi\omega$ interference phase, which is set to $163^\circ$
\cite{Achasov_3pi,SND_3pi}, and all the couplings involved in the reaction.
$C_{\rho\pi/\omega\pi}$ are complex coefficients that include
the uncertainty arising from 
the recurrences of the $\rho$ and $\omega$ mesons
and $\delta_{b_{\rho}}$ is the phase of the 
amplitude when the $\rho$ is the intermediate state.

%==============================================================================
% Bibliography
%==============================================================================

%%%%%%%%%%%%%%%%%%%%%%%%%%%%%%%%%%%%%%%%%%%%%%%%%%%%%%%%%%%%%%%%%%%%%%%%%%%%%%%
\end{document}